\documentclass{fic-l}
\def\H{\mathbf H}
\def\Q{\mathbb Q}
\def\k{\mathbf k}
\def\C{\mathbb C}
\def\bC{\mathbf C}
\def\R{\mathbb R}
\def\A{\mathbb A}
\def\bA{\mathbf A}
\def\B{\mathbf B}
\def\L{\mathbf L}
\def\V{\mathbf V}
\def\X{\mathbf X}
\def\G{\mathbf G}
\def\N{\mathbb N}
\def\U{\mathbf U}
\def\id{{\mbox{id}}}
\def\ugamma{{\underline{\gamma}}}

\newtheorem{theorem}{Theorem}[section]
\newtheorem{lemma}[theorem]{Lemma}
\newtheorem{reminder}[theorem]{Reminder}
\newtheorem{proposition}[theorem]{Proposition}
\newtheorem{def-lemma}[theorem]{Definition-Lemma}
\newtheorem{def-proposition}[theorem]{Definition-Proposition}
\newtheorem{corollary}[theorem]{Corollary}

\theoremstyle{definition}
\newtheorem{definition}[theorem]{Definition}                                                                              
\newtheorem{example}[theorem]{Example}
\newtheorem{notation}[theorem]{Notation}
\newtheorem{def-notation}[theorem]{Definitions and notation}

\theoremstyle{remark}
\newtheorem{remark}[theorem]{Remark}

\numberwithin{equation}{section}

\begin{document}

\title{Connes-Moscovici-Kreimer Hopf Algebras}

\author{Daniel Kastler}
\address{Centre de Physique Th\'eorique\\
CNRS - Luminy, Case 907\\
F-13288\\
Marseille Cedex 09, France\\
and\\
Universit\'e de la M\'editerrann\'ee}
\email{}

\dedicatory{Dedicated to Sergio Doplicher and John Roberts}

\begin{abstract}
These notes hopefully provide an aid to the comprehension of the Connes-Moscovici and Connes-Kreimer works, by isolating common mathematical features of the Connes-Moscovici, rooted trees, and Feynman-graph Hopf algebras (as a new special branch of the theory of Hopf algebras expected to become important).  We discuss in particular the dual Milnor-Moore situation.
\end{abstract}

\maketitle

\section{Characters and infinitesimal characters of Hopf algebras} \label{sec1}

In what follows $\mathbf{k}$ is a commutative field of zero characteristic 
(e.g. $\mathbf{k}= {\mathbb C, \mathbb R})$.

\begin{reminder}
{\em (the algebra\"{\i}c dual of Hopf 
algebras).} Let  $\H\ (m,\mathbf 1=e1,\Delta,\varepsilon,S)$ be a Hopf algebra over $\k$. Consider the algebraic 
dual $\H^* = (m_*, \mathbf 1_* = e_*1,\Delta_* = m^t,\varepsilon_* = e^t, S_*=S^t)$
of $\H$ endowed with the topology of simple convergence on $\H$.\footnote{the $\sigma(\H^*,\H)$-topology.}
The value of $\xi \in \H^*$ for $h \in \H$ is denoted by $<\xi,h>$.  Then:
\begin{enumerate}
\item[(i)] 
As the dual of the coalgebra $\H,\H^*$ is 
a genuine algebra with the following product\footnote{We note this 
product with a $*$ because it is the convolution product of End$_\C(\H,\C)$, cf. definition (2,1) in \cite{B}, namely, denoting the 
covalue $<\xi,h>$ by $\xi(h)$: $(\xi * \eta)(h) = [m_k(\xi \otimes \eta)\Delta](h)$,  
where we can omit $m_k$  owing to $\k\otimes\k\cong\k$.} $*$ and unit $\mathbf 1_*$:
\begin{eqnarray}\label{1.1}
\left\{
\begin{array}{l}
<\xi*\eta,h> = <\xi \otimes \eta,\Delta h>\\
<\mathbf 1_*,h> = \varepsilon(h)
\end{array}
\right. , \quad \xi,\eta \in \H^*,\ h \in \H.
\end{eqnarray}

\item[(ii)]  $\H^*$ is however not a Hopf algebra: we do have $S_* = S^t: \H^* \to \H^*$ and $\varepsilon_* =e^{t} : {\H}*\to \C$:
\begin{eqnarray}\label{1.2}
\left\{
\begin{array}{l}
<\xi,Sh> = <S_* \xi,h>\\
<\xi,\mathbf 1> = \varepsilon_*(\xi)
\end{array}
\right. , \quad \xi \in \H^*,\ h \in \H,
\end{eqnarray}
but in general $\Delta_{*}=m^{t}: \H^* \to (\H \otimes \H)^* \underset{\neq}{\supset} \H^* \otimes \H^*$.
\end{enumerate}
\end{reminder}

We now investigate two subsets of $\H^*$ which $\Delta_{*}$ maps 
into $\H^* \otimes \H^*$: The characters and infinitesimal 
characters of $\H$: 
\begin{definition}\label{d1.1}
 Let $\H(m,e,\Delta,\varepsilon)$ be a Hopf  $\k$-algebra.
\begin{enumerate}
\item[(i)] 
 a {\bf character of } $\H$ is a non vanishing $\k$-linear map 
$\chi: \H \rightarrow \k$ which is multiplicative:
\begin{eqnarray}\label{1.3}
<\chi,hk> = <\chi,h><\chi,k>, \qquad h,k \in \H
\end{eqnarray}
(note that, as a consequence, one has
\begin{eqnarray}\label{1.4}
<\chi,\mathbf 1> = 1,
\end{eqnarray}
indeed (\ref{1.3}) entails that $<\chi,\mathbf 1>[<\chi,\mathbf 1> - 1]$ vanishes, whence (\ref{1.4}) if $\chi$ does not vanish).
We denote by Char$\H$ the set of characters of $\H$. Note that 
the counit is a character of $\H$.
\item[(ii)] An {\bf infinitesimal character of} $\H$ is a $\k$-linear 
map $Z: \H\rightarrow\k$ fulfilling:\footnote{(\ref{1.5}) defines 
$Z$ as a derivation of the $\H$-bimodule $_\varepsilon\H_\varepsilon$ consisting of $\H$ with multiplication by $a \in \H$ from the left and from the right multiplication by $\varepsilon(a)$: 
$\left\{\begin{array}{l} h\cdot \alpha = \varepsilon(h)\alpha\\
\alpha \cdot h=\varepsilon(h)\alpha\end{array}\right. , \alpha \in \C,\ h \in \H.$)}
\begin{eqnarray}\label{1.5}
<Z,hk> = <Z,h> \varepsilon(k) + \varepsilon(h) <Z,k>, \qquad h,k \in \H
\end{eqnarray}
(note that, as a consequence, one has
\begin{eqnarray}\label{1.6}
<Z,\mathbf 1> = 0,
\end{eqnarray}
indeed one has 
$$<Z,\mathbf 1> = <Z,\mathbf 1 \cdot \mathbf 1>) = <Z,\mathbf 1> \varepsilon(\mathbf 1) + \varepsilon(\mathbf 1) <Z,\mathbf 1> = 2 <Z,\mathbf 1>;$$ and that $Z$ vanishes
 on the square $\H^{+2}$ of the augmentation 
ideal $\H^{+}= $ Ker$\varepsilon$). Note also that (\ref{1.5}) is the infinitesimal 
form of (\ref{1.3}) written for $\chi=\varepsilon+Z$ with $Z$ considered 
as small). 
\end{enumerate}
\end{definition}

We denote by $\partial$Char$\H$ the set of infinitesimal 
characters of $\H$.
\begin{proposition}
With the definitions in \ref{d1.1} we have that:
\begin{enumerate}
\item[(i)]
Equipped with the convolution product (\ref{1.1}), the 
unit $\mathbf 1_{*}$:
\begin{eqnarray}\label{1.1a}
\left\{
\begin{array}{l}
<\psi * \psi',h> = <\psi \otimes \psi',\Delta h>\\
<\mathbf 1_*,h> = \varepsilon(h)
\end{array}\right. , \quad \psi,\psi' \in \mbox{Char}\H,\ h \in \H.
\end{eqnarray}
the inverse:
\begin{eqnarray}\label{1.7}
\chi^{-1} = \chi \circ S, \qquad x \in \mbox{Char}\H,
\end{eqnarray}
and the $\sigma(\H^*,\H)$-topology, $\G = $Char$\H$ is 
a topological group, the {\bf group of characters of $\H$}.\footnote{in fact a subgroup of the group of invertibles of the algebra $\H^*$ with product $*$.}
\item[(ia)]
$\G$ is dually definable as the group of ``group-like 
elements" of $\H^*$:\footnote{``group-like elements" between quotation marks because $\H^*$ is not 
a Hopf algebra cf. (ii).}
\begin{eqnarray}\label{1.8}
\G = \{\psi \in \H^*, \psi \neq 0, \Delta_* \psi = \psi \otimes \psi\}
\end{eqnarray}
\item[(ii)] 
Equipped with the bracket:
\begin{eqnarray}\label{1.9}
{[}Z,Z'] = Z * Z' - Z' * Z, \qquad Z,Z' \in \partial\mbox{Char}\H,
\end{eqnarray}
where $*$ is the convolution product (\ref{1.1}):
\begin{eqnarray}\label{1.1b}
Z * Z' = <Z \otimes Z,\Delta h>, \qquad Z,Z' \in \partial\mbox{Char}\H,\ h \in \H,
\end{eqnarray}
$\L=\partial$Char$\H$ is a Lie algebra, the {\bf Lie 
algebra of infinitesimal characters of $\H$}.\footnote{in fact 
a Lie-subalgebra of the Lie algebra Lie$\H^*$ associated with the algebra $\H^*$.} 
\item[(iia)] $\L$ is dually definable as the Lie algebra of ``primitive 
elements" of $\H^*$:\footnote{``primitive elements" between quotation marks because $\H^*$ is not 
a Hopf algebra cf. (ii).} 
\begin{eqnarray}\label{1.10}
\L = \{Z \in \H^*,\Delta_* Z = Z \otimes {\mathbf 1}_* + {\mathbf 1}_*\otimes  Z\}.
\end{eqnarray}
\end{enumerate}
\end{proposition}

\begin{proof}
The product first line (\ref{1.1a}) is known to be associative. 
To assert that $\chi*\psi\in $Char$\H$ we need 
to check multiplicativity: now, for $h,k\in\H$:
\begin{eqnarray*}
<\chi * \psi,hk> &=& <\chi \otimes \psi, \Delta(hk)> = <\chi \otimes \psi,(\Delta h)(\Delta k)>\\
&=& <\chi \otimes \psi,\Delta h)><\chi \otimes \psi,\Delta k> = <\chi * \psi,h><\chi * \psi,k>
\end{eqnarray*}
where we used the multiplicativity of $\chi \otimes \psi$ due to the commutativity of $\k$:  
indeed, one has, for $x,y,x',y'$: 
\begin{eqnarray*}
<\chi \otimes \psi,(x\otimes y)(x'\otimes y')> &=& <\chi \otimes \psi,xx' \otimes yy'> = < \chi,xx'><\psi,yy'>\\
&=& <\chi,x><\chi,x'><\psi,y><\psi,y'> \\
&=& <\chi,x><\psi,y><\chi,x'><\psi,y'>\\
&=& <\chi \otimes \psi,x \otimes y><\chi \otimes \psi,(x' \otimes y')>
\end{eqnarray*}
Check that $\chi* {\mathbf 1}_*= {\mathbf 1}_* * \chi=\chi$: 
we have by the first line (\ref{1.1a}):
\begin{eqnarray*}
<\chi * {\mathbf 1}_*,h>&=& <\chi \otimes {\mathbf 1}_*,\Delta h> = <\chi \otimes {\mathbf 1}_*,h_{(1)} \otimes_{(2)}> \\
&=& <\chi,h_{(1)}><\varepsilon(h_{(2)})> = < \chi,\varepsilon(h_{(2)})h_{(1)}> \\
&=& <\chi,h>.\\
<{\mathbf 1}_* * \chi,h> &=& <{\mathbf 1}_* \otimes \chi,\Delta h> = <{\mathbf 1}_*\otimes \chi,h_{(1)}\otimes_{(2)} > \\
&=& < {\mathbf 1}_*,h_{(1)}><\chi,h_{(2)}>\\
&=& <\varepsilon(h_{(1)})><\chi,h_{(2)}> = <\chi,\varepsilon(h_{(1)})h_{(2)}> = < \chi,h>.
\end{eqnarray*}
Check of (\ref{1.7}) we have: 
\begin{eqnarray*}
{[}(\chi \circ S)*\chi](h) &=& [(\chi \circ S) \otimes \chi](\Delta h) = \chi (Sh_{(1)})\chi(h_{(2)}) = \chi(Sh_{(1)}a_{(2)})\\
&=& \chi(\varepsilon(h){\mathbf 1}) = \varepsilon(h)\chi({\mathbf 1}) = \varepsilon(h)\\
&&\\
{[}\chi  * (\chi \circ S)](h) &=& [\chi \otimes  (\chi \circ S)](\Delta h) = \chi(h_{(1)}) \chi(Sh_{(2)}) = \chi(h_{(1)})Sh_{(2)})\\
&=& \chi(\varepsilon(h){\mathbf 1}) = \varepsilon(h)\chi({\mathbf 1}) = \varepsilon(h).
\end{eqnarray*}
The facts that product and inverse are continuous in the $\sigma({\H}^*,\H)$-topology 
is clear.

\medskip

(ia)\ For the proof we write $\G = $ Char$\H$ and $\G' = \{\psi \in \H^*,\Delta_*\psi = \psi \otimes \psi\}$.

Check of $\G' \subset \G$: for $\psi \in \G'$, $h,h' \in \H$, we 
have: 
\begin{eqnarray*}
<\psi,h><\psi,h'> &=& <\psi \otimes \psi, h \otimes h'> = <\Delta_* \psi,h \otimes h'>\\
& =& <\psi,m(h \otimes h')> = <\psi,hh'>
\end{eqnarray*}

Check of $\G \subset \G'$: $\psi \in \G$ implies: 
\begin{eqnarray*}
<\psi,m(h \otimes h')> = < \psi,hh'> = <\psi,h><\psi,h'> = <\psi \otimes \psi,h \otimes h'>
\end{eqnarray*}
hence we have $<\psi,m(h \otimes h')> = <\Delta_*\psi,h \otimes h'>$ with 
$\Delta_*\psi = \psi \otimes \psi$.

The Lie algebra property of $\L$ can be checked directly from 
(\ref{1.5}), but will immediately result from (iia), which we now check. 

\medskip

(iia)\ For the proof we write $\L=\partial$Char$\H$ 
and $\L' = \{Z \in \H^*,\Delta_*Z = Z \otimes {\mathbf 1}_* + {\mathbf 1}_* \otimes Z\}$.

Check of $\L' \subset \L$: $Z \in \L'$ implies for $h,h' \in \H$: 
\begin{eqnarray*}
\lefteqn{<Z,h><{\mathbf 1}_*,h'> + <{\mathbf 1}_*,h><Z,h'>}\\
 && \qquad = <Z \otimes {\mathbf 1}_* + {\mathbf 1}_*
\otimes Z,h \otimes h'>\\
&& \qquad = <\Delta_*Z,h \otimes h'> = <Z,hh'> = <Z,m(h \otimes h')> = <Z,hh'>.
\end{eqnarray*}

Check of $\L \subset \L'$: for $Z \in \L$,  $h,h' \in \H$ one has, 
\begin{eqnarray*}
<Z,m(h \otimes h')> &=& <Z,hh'> = <Z,h><{\mathbf 1}_*,h'> + <{\mathbf 1}_*,h><Z,h'>\\
&=& <Z \otimes {\mathbf 1}_* + {\mathbf 1}_* \otimes Z,h \otimes h'>.
\end{eqnarray*}
hence we have $<Z,m(h \otimes h')>=<\Delta_*Z,h \otimes h'>$ with 
$\Delta_*\psi = Z \otimes {\mathbf 1}_* + {\mathbf 1}_* \otimes Z$.

Check of (ii): it is clear that $\L'$ is a Lie-subalgebra of 
the Lie algebra Lie$\H^*$.\end{proof}

\begin{def-lemma}\label{def1.3}
{\em Let $\H_{*}$ be 
the subalgebra of $\H^*$ generated by ${\mathbf 1}_*$ and $\L$.}
\begin{enumerate}
\item[(o)] One  has $\varepsilon_*(\xi * \eta) = \varepsilon_*(\xi)\varepsilon_*(\eta)$, $\xi,\eta \in \H^* \supset \H_*$, $\varepsilon_*({\mathbf 1}_*) = 1$
and $\varepsilon$ vanishes on $\L$.
 
\item[(i)] One has $S_{*}\H_{*} \subset \H_{*}$, $S_{*}\L \subset\L$, and
\begin{eqnarray}\label{1.11}
\left\{\begin{array}{l}
S_*(\xi * \eta) = (S_*\eta)*(S_*\xi),\xi,\ \eta \in \H^* \supset \H_*\\
S_*{\mathbf 1}_* = {\mathbf 1}_*.
\end{array}\right.
\end{eqnarray}

\item[(ii)] One has $\Delta_*\H_* \subset \H_* \otimes \H_*$ and $\Delta_*$
is multiplicative:
\begin{eqnarray}\label{1.12}
\Delta_*m_* = (m_* \otimes m_*)P_{12}(\Delta_* \otimes \Delta_*),
\end{eqnarray}
and cocommutative in the sense:
\begin{eqnarray}\label{1.13}
\Delta_* = P_{12}\Delta_*,
\end{eqnarray}
\end{enumerate}
\end{def-lemma}

\begin{proof}
 (o)\ One has $\varepsilon_{*}({\mathbf 1}_*)=<{\mathbf 1}_*,{\mathbf 1}>=1$, 
for $Z\in\L$, $\varepsilon_*(Z) = <Z,{\mathbf 1}>=0$. For 
$\xi \eta \in \H^*$ one has $\varepsilon_{*}(\xi *\eta) = <\xi * \eta, {\mathbf 1}> = < \xi \otimes \eta,\Delta {\mathbf 1}> = <\xi,{\mathbf 1}><\eta,{\mathbf 1}> = \varepsilon_*(\xi)\varepsilon_*(\eta)$.

(i)\ We first notice the property
\begin{eqnarray}\label{1.14}
\Delta_*S_* = (S_* \otimes S_*)P_{12}\Delta_*,
\end{eqnarray}
transposed of the known property $Sm=mP_{12}(S{\otimes}S)$.

Check of $S_{*}\L\subset\L$: for $Z\in\L$ one 
has $S_{*}Z\in\L$, indeed one has:
\begin{eqnarray*}
(\Delta_*S_*)Z &=& (S_* \otimes S_*)P_{12}\Delta_*Z\\
&=& (S_* \otimes S_*)P_{12}(Z \otimes {\mathbf 1} + {\mathbf 1} \otimes Z)\\
&=& S_* Z \otimes {\mathbf 1} + {\mathbf 1} \otimes S_*Z.
\end{eqnarray*}
It follows that $S_{*}\H_{*}\subset\H_{*}$. Now $S_{*}$ turns 
the unit ${\mathbf 1}_*$ into ${\mathbf 1}_*$. And, for $\xi,\eta \in \H^*$ and $h \in \H$, one has using the known property $(H\Delta)$: 
$\Delta S = (S \otimes S)P_{12}\Delta$:
\begin{eqnarray*}
<S_*(\xi * \eta),h> &=& <\xi * \eta,Sh> = < \xi \otimes \eta, \Delta Sh>\\
&=& <\xi \otimes \eta, (S \otimes S)P_{12}\Delta h> \\
&=& <S_* \eta \otimes S_* \xi,\Delta h> = < (S_*\eta ) * (S_* \xi),h>.
\end{eqnarray*}

(ii)\   (\ref{1.12}) reads $m^t\Delta^t = (\Delta^t \otimes \Delta^t)P_{12}(m^t \otimes m^t)$, transposed of the axiom $\Delta m = (m \otimes m)P_{12}(\Delta \otimes \Delta)$.

Check of the cocommutativity property (\ref{1.13}): holds obviously 
on the generators ${\mathbf 1}$ and $Z\in\L$. Now (\ref{1.13}) propagates 
multiplicatively: if $\xi,\eta \in \L$ fulfill $\Delta \xi = P_{12}\Delta\xi$, $\Delta\eta = P_{12}\Delta \eta$,
$\Delta(\xi\eta) = (\Delta \xi)(\Delta \eta) = (P_{12}\Delta\xi)(P_{12}\Delta \eta) = P_{12}(\Delta \xi)(\Delta \eta) = P_{12}\Delta(\xi \eta)$,
owing to multiplicatively of $P_{12}: P_{12}(a \otimes a')P_{12}(b \otimes b') = (a' \otimes a)(b'\otimes b) = a'a \otimes b'b = P_{12};(a \otimes a')(b \otimes b')]$.
\end{proof}

\begin{proposition}
$\ \H_{*}(m_{*},{\mathbf 1}_*,{\Delta}_{*},{\varepsilon}_{*},S_{*})\ $ 
is a cocommutative Hopf algebra.\footnote{In the graded 
connected case isomorphic to the enveloping algebra $\U(\L)$ 
of the Lie algebra $\L$ of its primitive elements (special case 
of the Milnor-Moore Theorem).}\\The Hopf algebras $\H$ and $\H_{*}$ are 
set in duality through the bilinear form $<\cdot,\cdot>$.\footnote{duality generally not separating $\H$ -- certainly not for $\H$ non-commutative., separating however in the case 
of CKM-Hopf algebras, cf. \ref{t3.5}.}
\end{proposition}
 
\begin{proof}
 $\H_{*}$ is a unital algebra with unit ${\mathbf 1}_*$ by 
definition. Using the terminology of Appendix A, we have 
that the coalgebra axioms $(C\Delta)$, $(C\varepsilon)$ 
for $\H_{*}$ result by transposition from the algebra axioms 
$(Am)$, resp. $(Ae)$ for $\H$. Bialgebra axioms for $\H_{*}$: we 
already met $(Bm)$ in (\ref{1.12}), $(Be)$ is the known fact that ${\mathbf 1_*} \in \G$, we have met $(B\varepsilon)$ and $(Be\varepsilon)$ in  Definition-Lemma \ref{def1.3}(o) above. Finally the axiom 
$(H)$ for $\H$ and $\H_{*}$ are transposed of each other. \end{proof}

\section{Graded Hopf algebras}\label{sec2}

\begin{notation}
With  $\H$ a $\C$-vector space direct 
sum of $\C$-vector subspaces:
\begin{eqnarray}\label{2.1}
\H = \otimes_{n \in \N} \H^n
\end{eqnarray}
we let ${\theta}_{z}$, $z{\in}\C$, resp.\ $Y$, be the $\C$-linear 
maps: $\H{\rightarrow}\H$ respectively specified by: 
\begin{eqnarray}\label{2.2}
\theta_z|_{\H^n} = e^{nz} \qquad \left(\Rightarrow \left\{\begin{array}{l}
\theta_{z+z'} = \theta_{z^o}\theta_z\\
\theta_z = e^{zY} \end{array}\right. ,\ z,z' \in \k \right),
\end{eqnarray}
\begin{eqnarray}\label{2.3}
Y_{\H^n} = n \qquad \left( \Leftrightarrow \frac{d}{dt} \theta_z|_{t=0}\right),
\end{eqnarray}
and set, for $a \in \H$ homogeneous:\footnote{$a{\in}H$ is 
{\bf homogeneous} whenever $a \in \bigcup_{n \in \N} \H^n$   
(${\Leftrightarrow}$ $a$ is an eigenvector of $Y$ -- then with eigenvalue 
${\partial}_{Y}(a)$).} 
\begin{eqnarray}\label{2.4}
\partial_Y(a) = n \mbox{ if } a \in \H^n , \qquad n \in \N.
\end{eqnarray}
\end{notation}

\begin{def-lemma}\label{def2.2}{\em
($\N$-graded algebras, coalgebras and bialgebras).

\medskip

(i)\  The $\C$-algebra $\H=\H(m,{\mathbf 1} = e(1))=\oplus_{n \in \N}\H^n$ is $\N$-{\bf graded} (with {\bf algebra}  $\N$-{\bf grading} $\oplus_{n \in \N}\H^n$) whenever one of the following {\it equivalent requirements} (\ref{2.5a}), (\ref{2.5b}), (\ref{2.5c}) {\it or} (\ref{2.5d}) prevails:
\begin{eqnarray}\label{2.5a}
\H^p \cdot \H^q \subset \H^{p+q}
 \qquad , p,q\in\N
\end{eqnarray}
\begin{eqnarray}\label{2.5b}
\left\{\begin{array}{l}
m \circ (\theta_z \otimes \theta_z) = \theta_{z}\circ m\\
\mbox{i.e.},\ \theta_z(ab) = (\theta_za)(\theta_zb),\ a,b \in \H
\end{array}\right. , \qquad z \in \C,
\end{eqnarray} 
(saying that ${\theta}_{z}$ is a one-parameter automorphism group 
of the algebra $\H(m,{\mathbf 1}=e(1))$, and implying ${\theta}_{z}({\mathbf 1})={\mathbf 1}$, 
$z{\in}\C$) 
\begin{eqnarray}\label{2.5c}
\left\{\begin{array}{l}
Y \circ m = m \circ (Y \otimes \id_{\H} + \id_{\H} \otimes Y)\\
\mbox{i.e.},\ Y(ab) = (Ya)b + a(Yb), \ a,b \in \H
\end{array}\right. .
\end{eqnarray}
(saying that $Y$ is a derivation of the algebra ${\H}(m, {\mathbf 1}=e(1))$, 
and implying $Y({\mathbf 1})=0$),
\begin{eqnarray}\label{2.5d}
\partial_Y(ab) = \partial_Y(a) + \partial_Y(b)
\end{eqnarray}
for $a,b{\in}\H$ homogeneous, which implies $ab{\in}{\H}$ homogeneous.

\medskip

(ii)\ The $\C$-coalgebra 
$\H = \H(\Delta,\varepsilon) = \oplus_{n \in \N}$ $\H^n$ is
 $N$-{\bf graded} (with {\bf coalgebra}  $N$-{\bf grading} ${\oplus}_{n\in\N}\H^{n}$) 
whenever one of the following {\it equivalent requirements} (\ref{2.6a}), (\ref{2.6b}), (\ref{2.6c}), or (\ref{2.6d}) prevails:\footnote{Note that (\ref{2.1}) obviously implies $\H \otimes \H = \sum_{i+j} \H^i \otimes \H^j$.}
\begin{eqnarray}\label{2.6a}
\Delta(\H^n) \subset \sum_{i+j = n} \H^i \otimes \H^j , \qquad n \in \N,
\end{eqnarray}
\begin{eqnarray}\label{2.6b}
(\theta_z \otimes \theta_z) \circ \Delta = \Delta \circ \theta_z, \qquad z \in \C,
\end{eqnarray}
(saying that ${\theta}_{z}$ is a one-parameter automorphism group 
of the coalgebra $\H(\Delta,\varepsilon)$, and implying 
$\varepsilon(\theta_z(a)) = \varepsilon(a)$, $a \in H(\Delta,\varepsilon)$)
\begin{eqnarray}\label{2.6c}
(Y \otimes \id_{\H} + \id_{\H} \otimes Y) \circ \Delta = \Delta \circ Y.
\end{eqnarray}
(saying that $Y$ is a coalgebra coderivation , and implying ${\epsilon}(Ya)=0$, 
$a{\in}{H}({\Delta},{\varepsilon})$) 
\begin{eqnarray}\label{2.6d}
\partial_Y(a) = \partial_Y(a_{(1)})+\partial_Y(a_{(2)}) \ \mbox{ if } \Delta a = a_{(1)} \otimes a_{(2)}, \ a \in \H,\mbox{ a homogeneous}.\nonumber\\
\end{eqnarray}

(iii)\ The $\C$-bialgebra $\H(m,\Delta,e,\varepsilon)=\oplus_{n\in\N}\H^{n}$ is $\N$-{\bf graded} (with {\bf bialgebra}  $\N$-{\bf grading} $\oplus_{n\in\N}\H^{n}$) 
whenever it is $\N$-graded both as an algebra and as a coalgebra. 
We say that $\H$ is $\N$-{\bf graded connected} whenever one has 
in addition:
\begin{eqnarray}\label{2.7}
\H^0 = \C{\mathbf 1}.
\end{eqnarray}

(iv)\ A Hopf algebra $\H(m,{\Delta}, e,{\varepsilon},S)$ is 
called $\N$-{\bf graded} whenever it is $\N$-graded as a bialgebra (cf. Definition-Lemma \ref{def2.2}(i)). $S$ {\it then automatically commutes with} 
$\theta_{z}$, $z{\in}\C$, {\it and with} $Y$:
\begin{eqnarray}\label{2.8}
\left\{\begin{array}{l}
\theta_z \circ S \circ \theta_z^{-1} = S\\
Y \circ S = S \circ Y
\end{array}\right. .
\end{eqnarray}
}\end{def-lemma}

For the $\N$-{\bf graded} algebra (resp. coalgebra, bialgebra) $\H$ we 
refer to $\theta_{z}$ as the {\bf automorphism group}, to $Y$ as 
the {\bf derivation} (resp. {\bf coderivation, biderivation}), 
and to $\partial_{Y}$ as the {\bf degree, of the} $N$-{\bf grading}. 

\begin{proof}
(i)\ (\ref{2.5a}) $\Leftrightarrow$ (\ref{2.5b}): $\Rightarrow$: 
it suffices to take $a{\in}{\H}^{p}$ and $b{\in}{\H}^{q}$, 
then $ab{\in}{\H}^{p+q}$, then: 
$$\theta_z(ab) = e^{z(p+q)}ab = (e^{zp}a)(e^{zq}b)b = (\theta_za)\theta_zb).$$
${\Leftarrow}$: for $a{\in}{\H}^{p}$ and $b{\in}{\H}^{q}$, $${\theta}_{z}(ab)=({\theta}_{z}a)({\theta}_{z}b)
=(e^{zp}a)(e^{zq}b)b=e^{z(p+q)}ab$$ 
thus $ab{\in}{\H}^{p+q}$.

(\ref{2.5a}) ${\Leftrightarrow}$ (\ref{2.5c}): ${\Rightarrow}$: it suffices to take
$a{\in}{\H}^{p}$ and $b{\in}{\H}^{q}$, then
$$Y(ab)=(p+q)ab=(pa)b+a(qb)=(Ya)b+a(Yb).$$ 
${\Leftarrow}$: for $a{\in}{\H}^{p}$ and $b{\in}{\H}^{q}$,
 $$Y(ab)=(Ya)b+a(Yb)=(pa)b+a(qb)=(p+q)ab $$
thus $ab{\in}{\H}^{p+q}$.

(\ref{2.5c}) ${\Leftrightarrow}$ (\ref{2.5d}): $Y(ab)={\partial}_{Y}(ab)ab$,whilst
$$(Ya)b+a(Yb)={\partial}_{Y}(a)ab+a{\partial}_{Y}(b)b
=[{\partial}_{Y}(a)+{\partial}_{Y}(b)]ab.$$ 

\medskip

(ii)\ (\ref{2.6a}) ${\Leftrightarrow}$ (\ref{2.6b}): ${\Rightarrow}$: it suffices 
to take $a{\in}{\H}^{p}$, then 
\begin{eqnarray*}
[({\theta}_{z}{\otimes}{\theta}_{z})\circ{\Delta}]a
&=& (\theta_z \otimes \theta_z)(a_{(1)} \otimes a_{(2)})\\
&=& (\theta_za_{(1)})\otimes (\theta_z a_{(2)}) = e^{p_1z}a_{(1)}\otimes e^{p_2z}a_{(2)}\\
&=& e^{pz}\Delta a = \Delta(e^{pz}a) = [\Delta \circ \theta_z]a.
\end{eqnarray*}
${\Leftarrow}$: for $a{\in}{\H}^{p}$, 
\begin{eqnarray*}
[(\theta_z \otimes \theta_z) \circ \Delta]a &=& (\theta_z \otimes \theta_z)(a_{(1)} \otimes a_{(2)})\\
&=& e^{p_1z}a_{(1)} \otimes e^{p_2z}a_{(2)}\\
&=& e^{(p_1+p_2)z}\Delta a = [\Delta \circ \theta_z]a = e^{pz}\Delta a,
\end{eqnarray*}
 implying $p_{1}+p_{2}=p$.\footnote{We chose the Sweedler basis 
$a_{(1)}{\otimes}a_{(2)}$ such that $a_{(1)}$, resp. the $a_{(2)}$, are 
linearly independent.} 

(\ref{2.6a}) ${\Leftrightarrow}$ (\ref{2.6c}): ${\Rightarrow}$: it suffices to take 
$a{\in}{\H}^{p}$, then:
\begin{eqnarray*}
[(Y \otimes \id_{\H} + \id_{\H} \otimes Y) \circ \Delta]a
&=& (Y \otimes \id_{\H} + \id_{\H} \otimes Y)(a_{(1)} \otimes a_{(2)})\\
&=& Ya_{(1)} \otimes a_{(2)} + a_{(1)} \otimes (Ya_{(2)})\\
&=& (p_1 + p_2)(a_{(1)} \otimes a_{(2)}) \\
&=& p \Delta(a) = \Delta(pa) = [\Delta \circ Y]a.
\end{eqnarray*}
${\Leftarrow}$: for $a{\in}{\H}^{p}$, 
\begin{eqnarray*}
[(Y \otimes \id_{\H} + \id_{\H} \otimes Y)\circ \Delta]a &=& (Y \otimes \id_{\H} + \id_{\H} \otimes Y)
(a_{(1)} \otimes a_{(2)})\\
&=& p_1a_{(1)} \otimes a_{(2)} + a_{(1)} \otimes p_2a_{(2)} \\
&=& (p_1 + p_2)(a_{(1)} \otimes a_{(2)})\\
&=& (p_1 + p_2)\Delta a = (\Delta \circ Y)a = p\Delta a
\end{eqnarray*}
implying $p_{1}+p_{2}=p$ as above.

(\ref{2.6c}) ${\Leftrightarrow}$ (\ref{2.6d}): ${\Delta}(Ya)={\Delta}[{\partial}_{Y}(a)a]=
{\partial}_{Y}(a){\Delta}a$,  whilst 
\begin{eqnarray*}
(Y \otimes \id_{\H} + \id_{\H} \otimes Y) \circ \Delta a &=& (Y \otimes \id_{\H} + \id_{\H} \otimes Y)(a_{(1)} \otimes a_{(2)})\\
&=& Ya_{(1)} \otimes a_{(2)} + a_{(1)} \otimes Ya_{(2)}\\
&=& \partial_Y(a_{(1)})a_{(1)} \otimes a_{(2)} + a_{(1)} \otimes \partial_Y(a_{(2)})a_{(2)}\\
& =& [\partial_Y(a_{(1)}) + \partial_Y(a_{(2)})] \Delta a.
\end{eqnarray*}

(iv)\ Check of the first line (\ref{2.8}): we check that ${\theta}_{z} \circ S \circ {\theta}_{z}^{-1}$ is 
a convolution inverse of $e{\varepsilon}$:
\begin{eqnarray*}
(\theta_z \circ S \circ \theta_z^{-1}) * \id_{\H} &=& m(\theta_z \circ S \circ \theta_z^{-1} \otimes \id_{\H})\Delta \\
&=& m(\theta_z \circ S \circ \theta_z^{-1} \otimes \id_{\H})(\theta_z \otimes \theta_z)\Delta \theta_z^{-1}\\
&=& m(\theta_z \otimes \id_{\H}(S \otimes \id_{\H})(\theta_z^{-1} \otimes \id_{\H})(\theta_z \otimes \theta_z)\Delta \theta_z^{-1}\\
&=& m(\theta_z \otimes \id_{\H})(S \otimes \id_{\H})(\id_{\H} \otimes \theta_z)\Delta \theta_z^{-1}\\
&=& m(\theta_z \otimes \id_{\H})(\id_{\H} \otimes \theta_z)(S \otimes \id_{\H})\Delta \theta_z^{-1}\\
&=& m(\theta_z \otimes \theta_z)(S \otimes \id_{\H})\Delta \theta_z^{-1} = \theta_z^{-1}m(S \otimes \id_{\H})\\
&=& \theta_z e \varepsilon \theta_z^{-1} = e\varepsilon,
\end{eqnarray*}
where we used (\ref{2.6b}), (\ref{2.5b}), ${\theta}_{z}({\mathbf 1})= {\mathbf 1}$, and ${\varepsilon}\circ{\theta}_{z}= {\varepsilon}$.

Check of the second line (\ref{2.8}): equate the derivations at 0 of 
${\theta}_{z}\circ S=S \circ \theta_{z}$. \end{proof}

In practice recognizing a biderivation often occurs by first 
establishing the derivation property, and afterwards the coderivation 
property, using 

\begin{lemma}\label{lem2.3}
 Assume that one knows that a 
bialgebra $\H$ is $\N$-graded as an algebra. For investigating 
the possible coderivation property of $Y$ one needs only look 
at the restriction of $Y$ on a set of generators. 
\end{lemma}

\begin{proof}
From
\begin{eqnarray*}
\left\{
\begin{array}{lll}
\partial_Y(a) = \partial_Y(a_{(1)}) + \partial_Y(a_{(2)}) & \mbox{if } & \Delta a = a_{(1)} \otimes a_{(2)}\\
\partial_Y(b) = \partial_Y(b_{(1)}) + \partial_Y(b_{(2)}) & \mbox{if } & \Delta a = a_{(1)} \otimes a_{(2)}
\end{array}\right. ,
\ a,b \in \H,
\end{eqnarray*}
$a,b$ homogeneous, we conclude
for the product $ab$ with coproduct $\Delta(ab) = a_{(1)} b_{(3)} \otimes a_{(2)}b_{(4)}$ that 
$\partial_Y(ab) = \partial_Y(a) + \partial_Y(b)$ equals
$$\partial_Y(a_{(1)}b_{(3)}) + \partial_Y(a_{(2)}b_{(4)}) = \partial_Y(a_{(1)}) + \partial_Y(b_{(3)}) + \partial_Y(a_{(2)}) + \partial_Y(b_{(4)}).$$

It is instructive to check instead the property (\ref{2.6c}), showing that
\begin{eqnarray*}
\lefteqn{(Y \otimes \id_{\H} + \id_{\H} \otimes Y)\Delta (ab)}\\
 &=& [(Y \otimes \id_{\H} + \id_{\H} \otimes Y)\Delta(a)]\Delta(b) + \Delta(a)[Y\otimes \id_{\H} + \id_{\H} \otimes Y)\Delta (b)]\\
&=& \Delta (Ya)\Delta(b) + \Delta(a) \Delta(Yb) = \Delta[(Ya)b + a(Yb)] = \Delta [Y(ab)].
\end{eqnarray*}
\end{proof}

 We next show that the algebra\"{\i}c dual of a $\N$-graded coalgebra 
is a $\N$-graded algebra.
\begin{proposition}\label{prop2.4}
{\em (the dual  N-grading).} With 
$\H(m,\Delta,e,\varepsilon)$, $\H = \oplus_{n \in \N} \H^n$, a $\N$-
graded bialgebra, let $\H^*$ be the algebra\"{\i}c dual 
of $\H$ endowed with the algebra-structure dual of the coalgebra-structure 
of $\H$, thus with product:\footnote{Here the convolution 
product is that of End$_{\C}(\H,\H^*)$, and $<\cdot, \cdot>$ is 
the duality bilinear form, cf. (\ref{1.1}) }
\begin{eqnarray}\label{2.9}
<\xi* \eta,h> = < \xi \otimes \eta, \Delta h>, \qquad \xi,\eta \in \H^*.
\end{eqnarray}
The definitions:
\begin{eqnarray}\label{2.10}
\left\{
\begin{array}{l}
<Y_*\xi,h>=<\xi,Yh>\\
<{\theta_*}_z\xi,h>=<\xi,\theta_zh>
\end{array}\right. , \ \xi \in \H^*, \ h \in \H,
\end{eqnarray}
then yield a  $\N$-grading $\H^*={\oplus}_{n\in\N}\H^{*n}$ 
of the dual algebra $\H^*$ where $\H^{*n} =\bigcap_{p \neq n} \H^{p \perp}$. 
One has the inclusions $Y_{*}$``$P(\H^*)$" ${\subset}$ ``$P({\H}^*)$" 
and ${\theta_*}_{z}$``$G({\H}^*)$" ${\subset}$ ``$G({\H}^*)$", 
$z{\in}\C$.\footnote{We define ``$P(\H^*)$" $= \{\xi \in \H^*, \Delta_*\xi = 
\xi\otimes{\mathbf 1} + {\mathbf 1} \otimes \xi\}$, ``$G({\H}^*)$" $= \{\xi \in \H^*, \Delta_*\xi = \xi\otimes  \xi\}$ (written with quotation marks because $\H^*$ is not a bona-fide 
coalgebra -- this notation becomes bona-fide in $\H_{*}$, cf. Section \ref{sec1}).}
$\H_*$ thus becomes a graded bialgebra.
\end{proposition}

\begin{proof}
 We check that one has $Y_{*}{\xi}=n{\xi}$ for 
${\xi}{\in}{\H}^{*n}$. With $\id_{\H}={\Sigma}_{n\in\N}P_{n}$ the 
decomposition of the unit associated to the direct sum  $\H=\oplus_{n\in\N}\H^{n}$, 
we have $Y={\Sigma}_{n{\in}\N}nP_{n}$, thus for ${\xi}{\in}{\H}^{*n}$ and 
$h{\in}{\H}$, 
$$<\xi,Yh> = <\xi,P_nYh> = < \xi,YP_nh> = n<\xi,P_nh> = n<\xi,h>=<Y_*\xi,h>,$$
 hence 
$Y_{*}{\xi}=n{\xi}$. The fact that (\ref{2.5c}) and (\ref{2.6c}) are 
transposed of each other then implies that $Y_{*}$ is a derivation 
of the algebra ${\H}^*$. The proof of ${\theta_{*}}_z\xi = e^{nz}\xi$ for ${\xi}{\in}{\H}^{*n}$ is analogous. Check of the two last claims:

\medskip

\noindent
-- for $\xi \in $ ``$P(\H^*)$": 
\begin{eqnarray*}
\Delta_* Y_*\xi &=& (\id \otimes Y_* + Y_*\otimes \id)\Delta_* \xi = (\id \otimes Y_* + Y_*\otimes \id)({\mathbf 1} \otimes \xi + \xi \otimes {\mathbf 1})\\
&=& ({\mathbf 1} \otimes Y_*\xi + Y_*\xi \otimes {\mathbf 1}).
\end{eqnarray*}

\noindent
-- for $\xi \in $ ``$G(\H^*)$" and $z \in \C$:
$$\Delta_* {\theta_{*}}_z\xi = ({\theta_{*}}_z \otimes {\theta_{*}}_z) \Delta_*\xi =
({\theta_{*}}_z \otimes {\theta_{*}}_z)(\xi \otimes \xi) = ({\theta_{*}}_z\xi \otimes {\theta_{*}}_z \xi).$$
\end{proof}

\begin{proposition}\label{p2.5}
{\em($\N$-graded connected bialgebras.)}
Let $$\H(m,{\Delta},e,{\varepsilon})={\oplus}_{n{\in}\N} 
\H^{n}$$ be a $\N$-graded connected bialgebra (i.e. $\H^{0}=\C{\mathbf 1}$) and 
let $\H^{+}= $ Ker${\varepsilon}$ (called the {\bf augmentation ideal} -- it is known that ${\H}=\C{\mathbf 1}{\oplus}{\H}^{+}$ cf. Proposition \ref{pA.2}). Then:

\medskip

\noindent
(i)\ one has:
\begin{eqnarray}\label{2.11}
\H^+ = \mbox{ Ker}\varepsilon = \oplus_{n \geq 1} \H^n
\end{eqnarray}

\noindent
(ii)\ one has the implication:
\begin{eqnarray}\label{2.12}
h \in \H^+, \ h \mbox{ homogeneous } \Rightarrow \Delta h = h \otimes {\mathbf 1} + {\mathbf 1} \otimes h + h' \otimes h^{\prime\prime},
\end{eqnarray}
with $h'{\otimes}h^{\prime\prime}$ a Sweedler-type sum of terms 
fulfilling either (and thus both) of the equivalent requirements:
\begin{enumerate}
\item[(a)] 
all $\delta_{Y}(h')$, ${\delta}_{Y}(h^{\prime\prime})$ are 
strictly smaller than ${\delta}_{Y}(h)$.\footnote{Note that 
the property $\delta_{Y}(h') + {\delta}_{Y}(h^{\prime\prime}) = \delta_Y(h)$
of the Hopf $\N$-grading merely entails that $\delta_{Y}(h')$, ${\delta}_{Y}(h^{\prime\prime}) \leq \delta_Y(h)$, whilst progressivity requires strict inequalities. Note also 
that applying ${\varepsilon}{\otimes}\id_{H}$ and $\id_{H}{\otimes}{\varepsilon}$ to 
both sides of (\ref{2.12}) one proves ${\varepsilon}(h')={\varepsilon}(h^{\prime\prime})=0$, 
hence that all $h'$ and $h^{\prime\prime}$ belong to ${\H}^{+}$.}
\item[(b)] all $\delta_{Y}(h')$, ${\delta}_{Y}(h^{\prime\prime})$ are 
strictly positive.
\end{enumerate}
\end{proposition}
(Equivalence of (a) and (b): since the bialgebra $\N$-grading entails 
that one has 
$$\delta_{Y}(h') + {\delta}_{Y}(h^{\prime\prime}) = \delta_Y(h), \quad
\delta_Y(h') = \delta_Y(h) - \delta_Y(h^{\prime\prime})$$
and
$$ \delta_Y(h^{\prime\prime}) = \delta_Y(h) - \delta_Y(h')$$
are simultaneously strictly positive). 

The consequence (\ref{2.12}) of $\N$-graded connectedness plays a prominent 
technical role both in the Connes-Moscovici index theory and 
in the Connes-Kreimer theory of renormalization.\footnote{In fact it already played the role in the early days of the theory 
of Hopf algebras, through the corresponding inductive definition 
of the antipode, cf. Proposition \ref{p2.6}.}
 We therefore cast the special name of {\bf progressiveness} for 
this property, \ref{p2.5}(ii) above then says that {\it a $\N$-graded 
connected bialgebra is progressive.}

\begin{proof}
 (i)\ Let $h{\in}{\H}^{n}$, $n{\geq}1$: by (\ref{2.6d}) 
we have 
\begin{eqnarray*}
 {\Delta}h&=&h_{n}{\otimes}1+{\Sigma}_{i=1,\ldots,{n-1}}{\Sigma}_{k}h_{n-i,k} {\otimes}h'_{i,k}+1{\otimes}h'_{n}\\
&& \qquad \mbox{(first subscript indicating the ${\partial}_{Y}$-grade).}
\end{eqnarray*}
Hence:
\begin{eqnarray*}
h&=& (\varepsilon \otimes \id)\Delta h = \varepsilon(h_n){\mathbf 1} + \Sigma_{i=1,\ldots,n-1}
\Sigma_k\varepsilon(h_{n-i,k})h'_{i,k} + h'_n\\
&& \qquad \mbox{whence $\varepsilon(h_n)=0$ and $h=h'_n$}\\
&=& (\id \otimes \varepsilon)\Delta h = h_n + \Sigma_{i=1,\ldots,n-1}\Sigma_kh_{n-i}\varepsilon(h'_{i,k})+\varepsilon(h'_n){\mathbf 1}\\
&&\qquad \mbox{whence $h=h_n$ and $\varepsilon(h'_n) = 0$},
\end{eqnarray*}
implying ${\varepsilon}(h)={\varepsilon}(h_{n})=0$. We proved that $\oplus_{n \geq 1} \H^n \subset \H^+ = $ Ker$\varepsilon$.
(In fact we found that:
\begin{eqnarray}\label{2.13}
\Delta h = h \otimes {\mathbf 1} + {\mathbf 1} \otimes h'_n + \Sigma_{i=1,\ldots,n-1} \Sigma_k h_{n-i,k} \otimes h'_{i,k}, \quad h \in \H^n,
\end{eqnarray}
first subscript indicating the ${\partial}_{Y}$-grade).

Conversely let $h{\in}{\H}^{+}= $ Ker${\varepsilon}$, with $h={\lambda}{\mathbf 1}+{\Sigma}_{n{\geq}1}h_{n}$, 
$h_{n}{\in}{\H}^{n}$. By what precedes we have $0={\varepsilon}(h)=\lambda$, 
hence $h={\Sigma}_{n{\geq}1}h_{n}$, proving that ${\H}^{+}{\subset}{\oplus}_{n{\geq}1}\H^{n}$.

\medskip

(ii)\ For $h{\in}{\H}^{n}$, (\ref{2.13}) implies (\ref{2.12}). However (\ref{2.12}) 
is linear w.r.t. $h$, hence holds for $h{\in}{\H}^{+}$ by (i).
\end{proof}

\begin{remark}\label{rem2.5a}
 (i)\ Progressiveness at large follows 
from progressiveness for generators of ${\H}^{+}$. Indeed:

\medskip

(ii)\ Let  $h,k{\in}{\H}^{+}$: if  (\ref{2.12}) holds for $h$ and $k$, 
it holds for the product  $hk$.
\end{remark}

\begin{proof}
(i)\ follows from (ii) and the multiplicativity 
of ${\Delta}$.

\medskip

(ii)\ assume that  $\Delta h = h \otimes {\mathbf 1} + {\mathbf 1} \otimes h + h' \otimes h^{\prime\prime}$ 
with all ${\delta}_{Y}(h')$, ${\delta}_{Y}(h^{\prime\prime})$ strictly 
positive, and $\Delta k = k \otimes {\mathbf 1} + {\mathbf 1} \otimes k +
k' \otimes k^{\prime\prime}$
with all  ${\delta}_{Y}(k')$, ${\delta}_{Y}(k^{\prime\prime})$ strictly 
positive. It follows that
\begin{eqnarray*}
\Delta(hk) &=& (hk)\otimes{\mathbf 1} + {\mathbf 1}\otimes(hk) + h \otimes k+ k \otimes h + hk' \otimes k^{\prime\prime} + k' \otimes h k^{\prime\prime}\\
&& + h'k \otimes h^{\prime\prime} + h' \otimes h^{\prime\prime} k + k'k \otimes k^{\prime\prime}
+ k' \otimes k^{\prime\prime}k + h'k' \otimes h^{\prime\prime}k^{\prime\prime},
\end{eqnarray*}
where all the tensor products but the two first have both factors 
of strictly positive degree.
\end{proof}

We now present two important consequences of progressiveness. 
The first is that it makes automatic the existence of a (recursively 
defined) antipode: 
\begin{proposition}\label{p2.6}
Assume that the $\N$-graded bialgebra ${\H}(m,{\Delta},e,{\varepsilon})= {\oplus}_{n{\in}\N}\H^{n}$ 
is  $\N$-graded connected (thus progressive). Then $\H$ is a Hopf algebra.
\end{proposition}

\begin{proof}
The requirements, for the $\C$-linear $S_{r}$, resp. $S_{l}$: 
${\H}{\rightarrow}{\H}$:
\begin{eqnarray}\label{2.14r}
&& \left\{\begin{array}{ll}
S_r{\mathbf 1} = {\mathbf 1}\\
S_rh = -h - h'S_rh^{\prime\prime}, & h \in \H^+ = \mbox{ker}\varepsilon
\end{array}\right.\nonumber\\
&&  \mbox{(saying that $\id_{\H}*S_r = m(\id_{\H}\otimes S_r)\Delta = e \varepsilon$)},
\end{eqnarray}
resp.
\begin{eqnarray}\label{2.14l}
&&\left\{\begin{array}{ll}
S_l{\mathbf 1} = {\mathbf 1}\\
S_lh = -h - (S_lh') h^{\prime\prime}, & h \in \H^+ = \mbox{ker}\varepsilon
\end{array}\right.\nonumber\\
&& \mbox{(saying that $\id_{\H}*S_l = m(S_l \otimes \id_{\H})\Delta = e \varepsilon$)},
\end{eqnarray}
determine $S_{r}$ and $S_{l}$ by induction w.r.t.\ the degree ${\delta}_{Y}$ as 
respective right- and left convolution inverses of id$_{{\H}}$ 
for the convolution product of End$_{\C}({\H},{\H})$: but these 
then coincide because $S_l*\id_{\H}*S_r = e\varepsilon * S_r = S_r = S_l * e \varepsilon = S_l$:
thus $S=S_{r}=S_{r}$ is a bilateral convolution inverse of id$_{{\H}}$, 
thus the (unique) antipode.

The second consequence of progressiveness of ${\H}$ is that, given 
$h{\in}{\H}$, there is a drastic limitation (depending on 
$h$) of non-vanishing values $<\xi,h>$, ${\xi}{\in}{\H}_{*}$. 
We recall that (cf. Section \ref{sec1}):

-- the algebra\"{\i}c dual $\H^*$ of $\H$ becomes a topological 
algebra if equipped with the $\sigma({\H}^*,{\H})$-topology 
and the algebra-structure ($m_{*}=^*$, ${\mathbf 1}_{*}=e_{*}(1)$) stemming from 
the coalgebra-structure of ${\H}$,

-- the subalgebra ${\H}_{*}$ generated in ${\H}^*$ by the unit ${\mathbf 1}_{*}$ and 
the Lie-algebra ${\L}$ of infinitesimal characters becomes by transposition 
a Hopf algebra ${\H}_{*}(m_{*}=^*)$, ${\mathbf 1}_{*}=e_{*}(1)$, ${\Delta}_{*},{\varepsilon}_{*},S_{*})$.
\end{proof}

\begin{lemma}\label{lem2.7} 
Let the bialgebra $\H(m,{\Delta},e,{\varepsilon})={\oplus}_{n{\in}\N}\H^{n}$ 
be $\N$-graded connected (thus progressive)\footnote{hence 
a Hopf algebra, cf. Proposition \ref{p2.6}} and let $P^{+}$ be the 
projection on $\H^+$ parallel to $\C{\mathbf 1}$ in $\H$.\footnote{
One has $H=\C{\mathbf 1} \oplus \H^{+}$, cf. (\ref{2.11}), and  $P^{+}=\id_{\H}-e{\varepsilon}$, 
cf. Proposition \ref{pA.2}. Correlatively  $P^{+ \otimes p}$ projects 
$\H^{{\otimes}p}$ onto $\H^{+{\otimes}p}$.} Then:

\medskip

(i)\ One has for $n{\in}\N$ the implication:
\begin{eqnarray}\label{2.15}
\partial_Y(h) \leq n \Rightarrow P^{+\otimes(n+1)}\Delta^{(n)}h = 0.
\end{eqnarray}

(ii)\ One has consequently for $n \in \N$ the implication:
\begin{eqnarray}\label{2.16}
\partial_Y(h) \leq n \ (\mbox{i.e. } h \in \oplus_{p \leq n} \H^p) &\Rightarrow& <Z_1* \cdots * Z_{n+1},h> = 0 \nonumber\\
&& \mbox{ for all } Z_1,\ldots,Z_{n+1} \in \L.
\end{eqnarray}
alternatively formulated as: \\
\begin{eqnarray}\label{2.16a}
Z_1,\ldots,Z_{n+1} \in \L \Rightarrow Z_1* \cdots * Z_{n+1} \in (\oplus_{p \leq n} \H^p)^\perp, \quad n \in \N,
\end{eqnarray}
where ${\perp}$ indicates the annihilator in the dual).
\end{lemma}

\begin{proof}
 (i)\  Since by (\ref{2.11}) a general element of ${\H}$ is 
of the form ${\lambda}{\mathbf 1}+h$ with $h{\in}{\H}^{+}$, and we have 
obviously  $P^{+{\otimes}(n+1)}\Delta^{(n)}{\mathbf 1}=0$, it is 
no restriction for the proof of the implication (\ref{2.15}) to assume 
that $h$ belongs to ${\H}^{+}$.The projection $P^{+{\otimes}(n+1)}$ applied 
to $\Delta^{(n)} h = h_{(1)} \otimes h_{(2)} \otimes \cdots \otimes h_{(n+1)}$
then suppresses all the products $h_{(1)} \otimes h_{(2)} \otimes \cdots \otimes h_{(n+1)}$
containing at least one factor in ${\H}^{0}=\C{\mathbf 1}$: the remaining products 
$h_{(1)} \otimes h_{(2)} \otimes \cdots \otimes h_{(n+1)}$ stem 
from the successive Sweedler-type summations $h'{\otimes}h^{\prime\prime}$ 
in (\ref{2.12}), thus have by progressiveness all their factors $h^{(p)}$ 
such that ${\delta}_{Y}(h^{(p)}{\geq} 1$  (observe, as noticed 
in footnote 6, that all $h', h^{\prime\prime}$ belong to ${\H}^{+}$). The 
equality 
$$\delta_Y(h^{(1)}) + \delta_Y(h^{(2)}) + \cdots + \delta_Y(h^{(n+1)}) = \delta_Y(h)$$
requires $n+1 \leq \delta_Y(h) \Leftrightarrow n < \delta_Y(h)$,
otherwise $P^{+{\otimes}(n+1)}\Delta^{(n)}h$ vanishes, whence 
(\ref{2.14r}) and (\ref{2.15}) for $h{\in}{\H}^{+}$, the extension to $h{\in}{\H}$ being 
trivial.

\medskip

(ii)\ The expression of the n-fold convolution product of  ${\H}^{{*}}$:
\begin{eqnarray}\label{2.17}
<\xi_1 * \cdots \xi_n,h>&=&<\xi_1 \otimes \cdots \otimes \xi_n,\Delta^{(n)}h>\nonumber\\
&=& < \xi_1 \otimes \cdots \otimes \xi_n,h_{(1)} \otimes h_{(2)} \otimes \cdots \otimes h_{(n+1)}>
\end{eqnarray}
yields for products of elements of ${\L}$ vanishing on ${\mathbf 1}$:
\begin{eqnarray}\label{2.18}
<Z_1 * \cdots * Z_{n+1},h> &=& < Z_1 \otimes \cdots \otimes Z_n, P^{+\otimes(n+1)}\Delta^{(n)}h>\nonumber\\
&& \qquad Z_1,\ldots,Z_{n+1} \in \L,\ h \in \H,
\end{eqnarray}
(\ref{2.16}) thus follows from (\ref{2.15}). Note that (\ref{2.16}) holds for $n=0$ 
(obvious directly - embodying this into our proof would require 
the convention ${\Delta}^{(0)}=\id_{\H}$).
\end{proof}

\begin{lemma}\label{lem2.8}
Assume that the $\N$-graded connected bialgebra $\H(m,{\Delta},e,{\varepsilon})={\oplus}_{n{\in}\N}\H^{n}$ 
has a countable vector space-basis $\{x_{i}\}_{i{\in}\N}$ consisting 
of homogeneous elements of non-decreasing degree, and let $\bar{\H}_*$ be 
the closure of $\H_{*}$ in $\H$. Then:

\medskip

(i)\ $\H^*$ is metrizable with the distance\footnote{
$d$ is a translation-invariant distance: one has $d(\xi,\eta) \geq 0$, $d(\xi,\eta) = d(\eta,\xi)$, $d(\xi,\eta) \leq d(\xi,\zeta) + d(\zeta,\eta)$, and $d(\xi+\zeta,\eta+\zeta) = d(\xi,\eta)$ 
for all $\eta,\zeta \in \bar{\H}_*$, and $d(\xi,\eta) = 0$, $\xi,\eta \in \bar{\H}_*$ implies
 $\xi=\eta$. These 
properties follow from the properties $f(\xi + \eta) \leq f(\xi) + f(\eta)$, $f(-\xi) = f(\xi)$ for 
all $\xi,\eta \in \bar{\H}_*$ plus the implication $f(\xi) = 0 \Rightarrow <\xi,x_i> = 0$ for 
all $i \in \N$. The equality of topologies arises as follows: 
the $V_{i}=\{{\xi}{\in}\bar{\H}_*;f({\xi}) < 2^{-i}\}$ are a basis 
of neighbourhoods of zero in the $d$-topology, with $U_{i}\{{\xi}{\in}\bar{\H}_*;|<{\xi},x_{i}>|<1\} \supseteq V_{i}.$} 
\begin{eqnarray}\label{2.19}
\left\{\begin{array}{l}
d(\xi,\eta) = f(\xi-\eta)\\
\mbox{where } f(\xi) = \Sigma_{i \in \N} 2^{-i} \inf(|<\xi,x_i>|,1) 
\end{array}\right. ,\ \xi,\eta \in \H^*,
\end{eqnarray}

(ii)\ $\bar{\H}_*$ is a closed topological subalgebra of $\H$.

\medskip

(iii)\ $\bar{\H}_*$ is stable under $S_{*}$ acting in $\H^{{*}}$.
\end{lemma}

\medskip

\noindent${\rightarrow}$  Note that at this point the question of whether $\Delta * \bar{\H}_* \subset  \bar{\H}_* \otimes  \bar{\H}_*$, and whether, if so, 
$$ \bar{\H}_*(m_*=^*, {\mathbf 1}_* = e_*(1),\Delta_*,\varepsilon_*,S_*)$$ is a Hopf algebra, remains open. 

\begin{proof}
 (i)\ is the known device for proving metrizability 
of a topological vector-space whose topology is specified by 
a countable set of semi-norms. One checks all the properties 
stated in footnote 9.

\medskip

(ii)\ follows from the known fact that the closure of a topological 
algebra is an algebra.

\medskip

(iii)\ Check of $S_{*} \bar{\H}_* \subset \H_*$: let ${\xi}{\in} \bar{\H}_*$, ${\xi}=\lim_{n=\infty}{\zeta}_{n}$ 
with ${\zeta}_{n}{\in}{\H}_{*}$, we have, for $h{\in}{\H}$:
$$
<S_*\zeta_n,h> \underset{n=\infty}{\rightarrow} , <\xi,Sh> = <S_*\xi,h>,$$
hence $S_*\zeta_n \underset{n=\infty}{\rightarrow} S_*\xi$, hence $S_*\xi \in \bar{\H}_*$.
\end{proof}

\begin{proposition}\label{p2.9}
Assuming that the $\N$-graded 
connected bialgebra \linebreak $\H(m,{\Delta},e,{\varepsilon})={\oplus}_{n{\in}\N}\H^{n}$ 
has a countable vector space-basis $\{x_{i}\}_{i\in\N}$ such 
that ${\partial}(x_{i}){\rightarrow}{\infty}$ for $n\rightarrow\infty$. Let $\bar{\H}_*$ be 
the closure of  $\H_{*}$ in  $\H$, we have that:

\medskip

(i)\ with $Z\in\L$, and $a_{i}{\in}\C$, $i{\in}\N$, 
the sequence ${\Omega}_{n}={\Sigma}_{i=1,\ldots,n}a_{i}Z^{*i}$ is 
a Cauchy-sequence of $\H^*$ which converges for $n{\rightarrow}{\infty}$ towards 
an element ${\Omega}{\in}\bar{\H}_*$.

\medskip

(ia)\ the same holds for multi-variable power-series of elements 
of $\L$.

\medskip

(ii)\ in particular the Mac-Laurin power-series of $e^{Z}$, 
$Z{\in}{\L}$, yields an element, denoted by $e^{Z}$, 
of $\bar{\H}_*$ which is a character of ${\H}$.
\end{proposition}

${\rightarrow}$ Note that at this point the question of whether the 
$e^{Z}$, $Z{\in}{\L}$, generate the group ${\G}$ of characters 
of ${\H}$ remains open. 

\begin{proof}
 (i)\ we have from (\ref{2.19}), with $i(n)=\inf\{i{\in}\N;{\partial}_{Y}(x_{i}){\geq}n \}$:
\begin{eqnarray*}
d(\Omega_n,\Omega_{n+p}) &=& \Sigma_{i\in\N} 2^{-i} \inf(|\Omega_{n+p}-\Omega_n,x_i>|,1)\\
&=& \Sigma_{i\in\N} 2^{-i} \inf(|<a_nZ^{*n}+ \cdots + a_{n+p}Z^{* n+p},x_i>|1)\\
&=& \Sigma_{i \geq i(n)} 2^{-i} \inf(|<a_nZ^{*n} + \cdots + a_{n+p} Z^{*n+p},x_i>|,1)\\
&\leq& \Sigma_{i \geq i(n)} 2^{-i} = 2^{-i(n) +1}
\end{eqnarray*}
whence our claim since  $i(n){\rightarrow}{\infty}$ for $n{\rightarrow}{\infty}$.

\medskip

(ia)\ the proof is the same up to notation.

\medskip

(ii)\ Particular case of what precedes. Now the exponential of 
an infinitesimal character, if it makes sense, is a character.
\end{proof}

\section{Connes-Moscovici-Kreimer Hopf algebras}\label{sec3}

We cast a special name for a type of Hopf algebras playing a 
prevalent role in index theory and in renormalized quantum field 
theory. In most of what follows $\C$ could be replaced by any commutative 
field $\k$ of characteristics 0.

\begin{definition}\label{d3.1}
$\H$ is a {\bf Connes-Moscovici-Kreimer Hopf algebra} (for shortness a {\bf CMK Hopf algebra}) whenever

\medskip

(i)\ $\H(m={\vee},{\mathbf 1}=e1)$ is the $\C$-algebra of (commutative) 
polynomials of countably many variables $x_{i}$: in other words 
$\H$ is the symmetric algebra over the (countably-dimensional) 
$\C$-vector space $\V$ with a specified basis $\X=\{x_{i}\}_{i{\in}\N}$: 
\begin{eqnarray}\label{3.1}
\H = \vee\V = \oplus_{p \in \N}\V^{\vee p} \qquad (\V^{\vee 0} = \C)
\end{eqnarray}
(we shall write indifferently $h{\vee}h'$ (symmetrized tensor 
product) or $hh'$ (product of polynomials) for the product of 
elements $h,h'{\in}{\H}$).\footnote{The algebra $\H$ is 
also definable as the universal envelope $\U(\V)$ of the 
Lie-algebra $\V$ with trivial brackets $[x_{i},x_{j}]$, $i,j{\in}\N$.}

The degree of monomials yields a $\N$-grading of the algebra $\H$ 
called its {\bf polynomial grading}.\footnote{An essential feature 
for the sequel of the polynomial grading is that $\H$ is generated 
in degrees 0 and 1. The upper index $p$ of the $p$th-grade summand 
$\V^{{\vee}^p}$ r.h. of (\ref{3.1}) indicates a $p$-fold power of 
the product of ${\H}$.}
We denote the corresponding degree by deg: deg$|_{\V}^{\vee p}=p$ 
(thus deg$x_{i}=1$).

\medskip

(ii)\ This algebra-structure is part of a Hopf structure ${\H}(m, 
{\mathbf 1}=e(1),{\Delta},{\varepsilon},S)$ such that 
\begin{eqnarray}\label{3.2}
\left\{\begin{array}{l}
\Delta x_i - x_i \otimes {\mathbf 1} - {\mathbf 1} \otimes x_i \in \H \otimes \V\\
\varepsilon(x_i) = 0
\end{array}\right. , \ i \in \N
\end{eqnarray}
(the second line implying that  Ker${\varepsilon}={\H}^{+}\supset{\oplus}_{n{\geq}1}V^{{\vee}n}$).

\medskip

(iii)\ In addition ${\H}$ is endowed with a Hopf  $\N$-grading 
\begin{eqnarray}\label{3.3}
\H = \oplus_{n \in \N} \H^n
\end{eqnarray}
(distinct from the (not Hopf) polynomial grading) for which all 
$x_{i}$ (thus all monomials) are homogeneous of strictly positive 
degree. Denoting respectively by $Y$, ${\delta}_{Y}$, and ${\theta}_{z}$, 
$z{\in}\C$, the corresponding biderivation:
\begin{eqnarray}\label{3.4a}
Y|_{\H^n}=n, \qquad n \in\N,
\end{eqnarray}
degree:
\begin{eqnarray}\label{3.4b}
\delta_Y(h) = n, \qquad h \in \H^n,\ n \in \N,
\end{eqnarray}
and automorphism group: 
\begin{eqnarray}\label{3.4c}
\theta_z|_{\H^n}=e^{nz}, \qquad n \in \N,
\end{eqnarray}
we thus have that
\begin{eqnarray}\label{3.5}
Yx_i= \delta_Y(x_i)x_i \ \mbox{ with }\ \delta_Y(x_i) \geq 1, \qquad n \in \N,
\end{eqnarray}
hence $Y$ and ${\theta}_{z}=e^{Yz}$, $z{\in}\C$, leave stable each 
of the subspaces $V^{{\vee}p}$, $p{\in}\N$ (the Hopf grading 
is {\bf subordinate} to the polynomial grading).

\medskip

As a consequence (cf. (\ref{2.11}) in Section \ref{sec2}), and using a previous 
remark, we have:
\begin{eqnarray}\label{3.8}
\oplus_{n \geq 1} \H^n = \H^+ = \mbox{ Ker}\varepsilon = \oplus_{n \geq 1} V^{\vee n}.
\end{eqnarray}

(iv)\ the $\N$-graded Hopf algebra ${\H}$ is $\N$-graded-connected: 
\begin{eqnarray}\label{3.6}
\H^0 = \C{\mathbf 1}.
\end{eqnarray}
thus progressive, i.e., such that, using a Sweedler-like notation, 
we have the implication:
\begin{eqnarray}\label{3.7}
h \in \H^+, \ h \mbox{ homogeneous } \Rightarrow \Delta h = h \otimes {\mathbf 1} + {\mathbf 1} \otimes h + h' \otimes h^{\prime\prime}
\end{eqnarray}
whereby all ${\delta}_{Y}(h')$, ${\delta}_{Y}(h^{\prime\prime})$ are strictly 
positive and strictly smaller than ${\delta}_{Y}(h)$.\footnote{The Hopf $\N$-grading is such that ${\delta}_{Y}(h')+{\delta}_{Y}(h^{\prime\prime})={\delta}_{Y}(h)$, 
entailing merely ${\delta}_{Y}(h')$, ${\delta}_{Y}(h^{\prime\prime}){\leq}{\delta}_{Y}(h^{\prime\prime})$. 
Note that we have that all $h',h^{\prime\prime}{\in}{\H}^{+}$ (cf. 
footnote 6 of Section \ref{sec2}).}
\end{definition}

\begin{remark}
We could replace in (ii) ``Hopf 
algebra" by ``bialgebra" since (\ref{3.7}) makes 
the existence of the antipode automatic, cf. Proposition \ref{p2.6}.
\end{remark}

\begin{proposition}\label{p3.3}
Let $\H$ be a CMK 
Hopf algebra. The Lie algebra $\L= {\partial}$Char$\H$ is 
in linear bijection with the algebra\"{\i}c dual of $\V$: 
a $Z{\in}{\L}$ is of the type: 
\begin{eqnarray}\label{3.9}
Z = \left\{\begin{array}{lll}
0 & \mbox{on} & \C{\mathbf 1}\\
\phi & \mbox{on} & \V\\
0 & \mbox{on} & \oplus_{n \geq 2} \V^{\vee n} = \H^{+2}
\end{array}\right. , \ \phi \mbox{ a linear form on $\V$},
\end{eqnarray}
all elements of $\L$ being obtained in this way (we then 
write $Z{\leftrightarrow}{\phi}$).
\end{proposition}

\begin{proof} 
Equality ${\oplus}_{n{\geq}1}\V^{{\vee}n}={\H}^{+}$: 
Since ${\oplus}_{n{\geq}1}\V^{{\vee}n}={\H}^{+}$, cf. 
(\ref{3.8}), the inclusion ${\supset}$ is obvious, the inclusion ${\subset}$ following 
from the fact that the algebra ${\H}$ is generated in polynomial 
grade 0 and 1. The definition property 
$$Z \in \L= {\partial}\mbox{Char}(\H) {\Leftrightarrow} 
<Z,hh'> = <Z,h>\varepsilon(h') + \varepsilon(h)<Z,h>,$$
$h,h' \in \H$ entails for $h=h'=1$ that $<Z,{\mathbf 1}>=0$, and 
that $<Z,hh'>=0$ for $h,h'{\in}{\H}^{+}=$ Ker${\varepsilon}$. 
Conversely for $Z$ as in (\ref{3.9}) one has $Z \in \L= {\partial}$Char$(\H)$
indeed:

-- $<Z,{\mathbf 1}{\mathbf 1}> = <Z,{\mathbf 1}> = 0$ whilst $\varepsilon({\mathbf 1})<Z,{\mathbf 1}> +<Z,{\mathbf 1}>\varepsilon({\mathbf 1}) = 0$,

-- for $h \in \H^1$: $<Z,{\mathbf 1} h> = <Z,h> = \phi(h)$ whilst $\varepsilon({\mathbf 1})<Z,h> + <Z,{\mathbf 1}>\varepsilon(h)  = 1\phi(h) = \phi(h)$,

-- for $h,h' \in \oplus_{n \geq 1}\H^n = \H^+$: $hh' \in \H^{+2}$ hence $<Z,hh'> = 0$ whilst $\varepsilon(h) <Z,h'> + <Z,h>\varepsilon(h') = 0$ since $\varepsilon(h')$ and $\varepsilon(h)$ both vanish.
\end{proof}

The next result, a partial $n$-fold generalization of Proposition \ref{p3.3}, 
is a tool for proving Theorem \ref{t3.5}

\begin{corollary}\label{c3.4}
 (i)\ Let $\H$ be a CMK Hopf algebra with $Z_{i}{\leftrightarrow}{\phi}_{i}{\in}{\L}$, 
$i=1,\ldots,n$, and $x_{i_p}{\in}{\V}$, $p=1,\ldots,n$, $n{\geq}1$, we have that:
\begin{eqnarray}\label{3.10n}
<Z_1 * \cdots * Z_n, x_{i_1}\ldots x_{i_n}> = \Sigma_\sigma \phi_{\sigma_1}(x_{i_l}) 
 \cdots \phi_{\sigma_n}(x_{i_n})
\end{eqnarray}
(the  sum ${\Sigma}_{\sigma}$ is over the set ${\Pi}_{n}$ of 
permutations of the $n$ first integers), consequently $\V^{{\vee}n}$ is 
separated by  the $n$-fold product $\L * \cdots * \L$: 
one has the implication:
\begin{eqnarray}\label{3.11}
h \in \V^{\vee n},\quad <Z_1 * \cdots * Z_n,h> = 0 \mbox{ for all } Z_1,\ldots,Z_n \in \L \Rightarrow h=0.
\end{eqnarray}

\medskip

(ii)\ We have, for $Z_1,\ldots,Z_n$, $n \in \N$, $n \geq 1$:
\begin{eqnarray}\label{3.12n}
<Z_1 * \cdots * Z_n, x_1\ldots x_m > = 0, \quad x_{i_p} \in \V, \ p = 1,\ldots,m, \ m > n \geq 1,
\end{eqnarray}
(in other terms:
\begin{eqnarray}\label{3.12a}
Z_1 * \cdots * Z_n|_{\H^{+m}} = 0, \qquad m > n \geq 1).
\end{eqnarray}
\end{corollary}
(Warning: (\ref{3.12n}) with $1{\leq}m<n$ generally does not 
hold: for instance, for $h{\in}{\H}^+$ one has
$<Z_1*Z_2,h>=<Z_1,h'>\phi_2(h^{\prime\prime})$.)

\begin{proof}
 (i)\ Before proving (\ref{3.10n}) and (\ref{3.12n}) for 
general $n$, we look for orientation at the cases $n=1,2,3$, and 
4.

\medskip

\noindent
{\bf Case $n=1$}: (\ref{3.10n}) and (\ref{3.12n})  amount to the second, 
resp. third line (\ref{3.9}).

\medskip

\noindent
{\bf Case $n=2$}: We first prove
\begin{eqnarray}\label{3.102}
&& \left\{\begin{array}{l}
<Z_1 * Z_2,x_{i_1}x_{i_2}>\\
= <Z_1\otimes Z_2,\Delta(x_{i_1}x_{i_2})>
\end{array}\right.\nonumber\\
&& \qquad = \phi_1(x_{i_1})\phi_2(x_{i_2})+ \phi_2(x_{i_1})\phi_1(x_{i_2}),\ Z_1 \leftrightarrow \phi_1,\ Z_2 \leftrightarrow \phi_2 \in \L.
\end{eqnarray}

In order to simultaneously prepare the next steps we compute 
$\Delta(\lambda_1\lambda_2) = \Delta(\lambda_1)\Delta(\lambda_2)$ for general $\lambda_1,\lambda_2 \in \H^+$: computation 
indicated by the table (with the terms of $\Delta(\lambda_1)$) 
in the first column, the terms of ${\Delta}({\lambda}_{2}$) in 
the first line, and the products at the intersection of the other 
lines and columns.)

\clearpage
\begin{center}
{Table $T_2$ : Computation of $\Delta(\lambda_1\lambda_2)$}
\end{center}
\begin{center}
\begin{tabular}{cccc}
& $\lambda_2 \otimes {\mathbf 1}$ & ${\mathbf 1} \otimes \lambda_2$ & $\lambda'_2 \otimes \lambda^{\prime\prime}_2$\\
&&&\\
$\lambda_1 \otimes {\mathbf 1}$ & \$$\lambda_1\lambda_2 \otimes {\mathbf 1}$ & $\lambda_1 \otimes \lambda_2$ & \$$\lambda_1\lambda'_2 \otimes \lambda^{\prime\prime}_2$\\
${\mathbf 1} \otimes \lambda_1$ & $\lambda_2 \otimes \lambda_1$ & \#${\mathbf 1} \otimes \lambda_1\lambda_2$ & \#$\lambda'_2 \otimes \lambda_1\lambda^{\prime\prime}_2$\\
$\lambda'_1 \otimes \lambda^{\prime\prime}_1$ & \$$\lambda'_1\lambda_2 \otimes \lambda^{\prime\prime}_1$ & \#$\lambda'_1 \otimes \lambda^{\prime\prime}_1\lambda_2$ & \#$\lambda'_1\lambda'_2 \otimes \lambda^{\prime\prime}_1\lambda^{\prime\prime}_2$
\end{tabular}
\end{center}
yields 9 tensor products:

-- the $4=4(2-1)$ indicated by \# belong to ${\H}{\otimes}{\H}^{+2}$, 
thus vanish under $<Z_{1}{\otimes}Z_{2}, \cdot >$,\footnote{in fact, as we will see, these can be discarded for the whole sequel.}

-- the $3=2+1$ indicated by \$ belong to ${\H}^{+2}{\otimes}{\H}$, 
thus also vanish under $<Z_{1}{\otimes}Z_{2}, \cdot>$,

-- the remaining 2: ${\lambda}_{1}{\otimes}{\lambda}_{2}$ and ${\lambda}_{2}{\otimes}{\lambda}_{1}$, yield 
(\ref{3.102}) upon making ${\lambda}_{1}=x_{i_1}$, ${\lambda}_{2}=x_{i_2}$.

We next prove
\begin{eqnarray}\label{3.122}
\left\{\begin{array}{l}
<Z_1 * Z_2,x_{i_1}\ldots x_{i_m}>\\
= < Z_1 \otimes Z_2,\Delta(x_{i_1}\ldots x_{i_m})>
\end{array}\right.
= 0 \mbox{ if } m > 2.
\end{eqnarray}

Since $x_{i_1} \ldots x_{i_m}$, $m> 2$, is of the form $\lambda_1\lambda_2\lambda_3$ with $\lambda_1,\lambda_2,\lambda_3 \in \H^+$, we 
compute the relevant part of $\Delta(\lambda_1,\lambda_2,\lambda_3) = \Delta(\lambda_1,\lambda_2)\Delta\lambda_3$ whereby the \#-terms of $\Delta(\lambda_1\lambda_2)$ can be discarded since multiplications by $\Delta\lambda_3$ leave 
${\H}{\otimes}{\H}^{+2}$ invariant: we thus have the table:

\begin{center}
Table $T_3$: Computation of $\Delta(\lambda_1\lambda_2\lambda_3)$
\end{center}
\begin{center}
\begin{tabular}{cccc}
& $\lambda_3 \otimes {\mathbf 1}$ & ${\mathbf 1} \otimes \lambda_3$ & $\lambda'_3 \otimes \lambda^{\prime\prime}_3$\\
&&&\\
$\lambda_1\lambda_2\otimes {\mathbf 1}$ & \$$\lambda_1\lambda_2\lambda_3\otimes {\mathbf 1}$ & $\lambda_1\lambda_2\otimes \lambda_3$ & \$$\lambda_1\lambda_2\lambda'_3\otimes \lambda^{\prime\prime}_3$\\
$\lambda_1 \otimes \lambda_2$ & $\lambda_1\lambda_3 \otimes \lambda_2$ & \#$\lambda_1 \otimes \lambda_2\lambda_3$ & \#$\lambda_1\lambda'_3 \otimes \lambda_2 \lambda^{\prime\prime}_3$\\
$\lambda_1\lambda'_2 \otimes\lambda^{\prime\prime}_2$ &  \$$\lambda_1\lambda'_2\lambda_3 \otimes \lambda^{\prime\prime}_2$ & \#$\lambda_1\lambda'_2 \otimes \lambda^{\prime\prime}_2\lambda_3$ & \#$\lambda_1\lambda'_2\lambda'_3 \otimes \lambda^{\prime\prime}_2\lambda^{\prime\prime}_3$\\
$\lambda_2 \otimes \lambda_1$ & $\lambda_2\lambda_3 \otimes \lambda_1$ & \#$\lambda_2 \otimes \lambda_1\lambda_3$ & \#$\lambda_2\lambda'_3 \otimes \lambda_1\lambda^{\prime\prime}_3$\\
$\lambda'_1\lambda_2 \otimes \lambda^{\prime\prime}_1$ & \$$\lambda'_1\lambda_2\lambda_3 \otimes \lambda^{\prime\prime}_1$ & \#$\lambda'_1\lambda_2 \otimes \lambda^{\prime\prime}_1\lambda_3$ & \#$\lambda'_1\lambda_2\lambda'_3 \otimes \lambda^{\prime\prime}_1\lambda^{\prime\prime}_3$
\end{tabular}
\end{center}
yielding 15 tensor products, all in $\H^{+2}{\otimes}{\H}$ or $\H{\otimes}{\H}^{+2}$, 
thus vanishing under $Z_{1}{\otimes}Z_{2}$.

\medskip

\noindent
{\bf Case $n=3$}: We first prove
\begin{eqnarray}\label{3.113}
&& \left\{\begin{array}{l}
<Z_1 * Z_2 * Z_3,x_{i_1}x_{i_2}x_{i_3}>\\
= <Z_1 \otimes Z_2 \otimes Z_3,(\Delta \otimes\id)\Delta (x_{i_1}x_{i_2}x_{i_3})>
\end{array}\right.\nonumber\\
&& \qquad = \Sigma_\sigma \phi_{\sigma_1}(x_{i_1})\phi_{\sigma_2}(x_{i_2})\phi_{\sigma_3}(x_{i_3}),\ Z_i \leftrightarrow \phi_i \in \L,\ i = 1,2,3.
\end{eqnarray}

We need to classify the tensor products in Table $T_3$: They consist 
of:

-- the $8=4(3-1)$ indicated by \# belong to ${\H}{\otimes}{\H}^{+2}$, 
thus vanish under $Z_{1}{\otimes}Z_{2}{\otimes}Z_{3}$$^{23}$,

-- the $4=3+1$ indicated by \$ which belong to ${\H}^{+3}{\otimes}{\H}$, 
turned by ${\Delta}{\otimes}\id$ into $({\Delta}{\H}^{+3}){\otimes}{\H}$ 
whose first factor  vanishes under $Z_{1}{\otimes}Z_{2}$, 

-- the remaining 3: $\lambda_{\sigma_1},\lambda_{\sigma_2}\otimes\lambda_{\sigma_3}$, $\sigma$ in the set ${\Pi}^0_3$ of cyclic permutations of $\{1,2,3\}$, 
whose sum is turned by ${\Delta}{\otimes}\id$ into ${\Sigma}_{\sigma\in\Pi^0_3}{\Delta}(\lambda_{\sigma_1}\lambda_{\sigma_2})\otimes \lambda_{\sigma_3}$, 
by Table $T_2$ equal modulo $\H \otimes \H^{+2} \otimes \H + \H^{+2} \otimes \H \otimes \H$ to $\Sigma_{\sigma\in \Pi^0_3} (\lambda_{\sigma_1} \otimes \lambda_{\sigma_2} + \lambda_{\sigma_2} \otimes \lambda_{\sigma_1}) \otimes \lambda_{\sigma_3}$ (in other terms $\Sigma_{\sigma\in\Pi_\sigma} \lambda_{\sigma_1} \otimes \lambda_{\sigma_2} \otimes \lambda_{\sigma_3}$:
we proved (\ref{3.113}).

We next prove
\begin{eqnarray}\label{3.123}
\left\{\begin{array}{l}
<Z_1*Z_2*Z_3,x_{i_1}\ldots x_{i_m}>\\
= <Z_1 \otimes Z_2 \otimes Z_3,(\Delta\otimes\id)(x_{i_1}\ldots x_{i_m})>
\end{array}\right. = 0 \mbox{ if } m > 3.
\end{eqnarray}

Since $x_{i_1} \ldots x_{i_m}$, $m>3$, is of the form 
$\lambda_1\lambda_2\lambda_3\lambda_4$ with $\lambda_1,\lambda_2,\lambda_3,\lambda_4 \in \H^+$, we compute the relevant part of $\Delta(\lambda_1\lambda_2\lambda_3,\lambda_4) = \Delta(\lambda_1\lambda_2\lambda_3)\Delta\lambda_4$
whereby the \#-terms of $\Delta(\lambda_1\lambda_2\lambda_3)$ can be discarded since multiplications by $\Delta\lambda_4$ leave $\H\otimes \H^{+2}$
 invariant: we thus have the table:

\begin{center}
Table $T_4$ : Computation of $\Delta(\lambda_1\lambda_2\lambda_3\lambda_4)$ 
\end{center}
\begin{center}
\begin{tabular}{cccc}
& $\lambda_4 \otimes {\mathbf 1}$ & ${\mathbf 1} \otimes \lambda_4$ & $\lambda'_4 \otimes \lambda^{\prime\prime}_4$\\
&&&\\
$\lambda_1\lambda_2\lambda_3 \otimes {\mathbf 1}$ & \$$\lambda_1\lambda_2\lambda_3\lambda_4 \otimes {\mathbf 1}$ & $\lambda_1\lambda_2\lambda_3 \otimes \lambda_4$ & \$$\lambda_1\lambda_2\lambda_3\lambda'_4 \otimes \lambda^{\prime\prime}_4$\\
$\lambda_1\lambda_2 \otimes \lambda_3$ & $\lambda_2\lambda_2\lambda_4 \otimes \lambda_3$ & \#$\lambda_1\lambda_2 \otimes \lambda_3\lambda_4$ & \#$\lambda_1\lambda_2\lambda'_4 \otimes \lambda_3\lambda^{\prime\prime}_4$\\
$\lambda_1\lambda_2\lambda'_3 \otimes \lambda^{\prime\prime}_3$ & \$$\lambda_1\lambda_2\lambda'_3 \lambda_4 \otimes \lambda^{\prime\prime}_3$ & \#$\lambda_1\lambda_2\lambda'_3 \otimes \lambda_4\lambda^{\prime\prime}_3$ & \#$\lambda_1\lambda_2\lambda'_3\lambda'_4 \otimes \lambda^{\prime\prime}_3\lambda^{\prime\prime}_4$\\
$\lambda_1\lambda_3 \otimes \lambda_2$ & $\lambda_1\lambda_3\lambda_4 \otimes \lambda_2$ & \#$\lambda_1\lambda_3 \otimes \lambda_2\lambda_4$ & \#$\lambda_1\lambda_3\lambda'_4 \otimes \lambda_2\lambda^{\prime\prime}_4$\\
$\lambda_1\lambda'_2\lambda_3 \otimes \lambda^{\prime\prime}_2$ & \$$\lambda_1\lambda'_2\lambda_3\lambda_4 \otimes \lambda^{\prime\prime}_2$ & \#$\lambda_1\lambda'_2\lambda_3 \otimes \lambda^{\prime\prime}_2\lambda_4$ & \#$\lambda_1\lambda'_2\lambda_3\lambda'_4 \otimes \lambda^{\prime\prime}_2\lambda^{\prime\prime}_4$\\
$\lambda_2\lambda_3 \otimes \lambda_1$ & $\lambda_2\lambda_3\lambda_4 \otimes \lambda_1$ & \#$\lambda_2\lambda_3 \otimes \lambda_1\lambda_4$ & \#$\lambda_2\lambda_3\lambda'_4 \otimes \lambda_1\lambda^{\prime\prime}_4$\\
$\lambda'_1\lambda_2\lambda_3 \otimes \lambda^{\prime\prime}_1$ & \$$\lambda'_1\lambda_2\lambda_3\lambda_4 \otimes \lambda^{\prime\prime}_1$ &
\#$\lambda'_1\lambda_2\lambda_3 \otimes \lambda^{\prime\prime}_1\lambda_4$ &
\#$\lambda_1'\lambda_2\lambda_3\lambda'_4 \otimes \lambda^{\prime\prime}_1\lambda^{\prime\prime}_4$
\end{tabular}
\end{center}
yielding 21 tensor products, all in ${\H}^{+3}{\otimes}{\H}$ or 
${\H}^{+2}{\otimes}{\H}^{+2}$, 
thus vanishing under $Z_{1}{\otimes}Z_{2}{\otimes}Z_{3}$, because 
$\Delta{\H}^{+3}$ vanishes under $Z_{1}{\otimes}Z_{2}$ by (\ref{3.122}).

A more detailed analysis of this table registrates, amongst its 
tensor products:

-- the $12=4(4-1)$ indicated by \# which belong to ${\H}{\otimes}{\H}^{+^2}$ 
turned by $(\Delta \otimes \id \otimes \id)(\Delta \otimes \id)$
into $\H \otimes \H \otimes \H \otimes \H^{+2}$ vanishing under 
$Z_{1}{\otimes}Z_{2}{\otimes}Z_{3}{\otimes}Z_{4}$$^{23}$, 

-- the $5=4+1$ indicated by \$ which belong to ${\H}^{+4}{\otimes}{\H}$ 
turned by $(\Delta \otimes \id \otimes \id)(\Delta \otimes \id)$
into $(\Delta \otimes \id)(\Delta\H^{+4}) \otimes \H$
vanishing under $Z_{1}{\otimes}Z_{2}{\otimes}Z_{3}{\otimes}Z_{4}$ because 
$(\Delta \otimes \id)\Delta\H^{+4}$ vanishes under 
$Z_{1}{\otimes}Z_{2}{\otimes}Z_{3}$: indeed $\Delta \otimes \id$
acting on the tensor products in $T_4$ contained in ${\H}^{+3}{\otimes}{\H}$, 
and ${\H}^{+3}$  vanishes under $Z_{1}{\otimes}Z_{2}$ by (\ref{3.122}).

-- the remaining 4: $\lambda_{\sigma_1}\lambda_{\sigma_2}\lambda_{\sigma_3}\otimes \lambda_{\sigma_4}$, $\sigma$ in the set ${\Pi}^0_4$ of cyclic permutations of $\{1,2,3,4\}$. 

We now reexamine (\ref{3.113}). We have, ${\cong}$ denoting equality 
up to negligible terms:
$$\Delta(\lambda_1\lambda_2\lambda_3) \cong \Sigma_{\tau \in \Pi^0_3} \lambda_{\tau_1}\lambda_{\tau_2} \otimes \lambda_{\tau_3},$$
 hence $(\Delta \otimes \id)\Delta(\lambda_1\lambda_2\lambda_3) \cong
\Sigma_{\tau \in \Pi^0_3} \Delta(\lambda_{\tau_1}\lambda_{\tau_2}) \otimes \lambda_{\tau_3}$,
 hence 
\begin{eqnarray*}
\lefteqn{
<Z_1 * Z_2 * Z_3 *,x_{i_1}x_{i_2}x_{i_3}> }\\
&=&
<Z_1 \otimes Z_2 \otimes Z_3,(\Delta \otimes \id)\Delta(x_{i_1}x_{i_2}x_{i_3})>\\
&=& \Sigma_{\tau \in \Pi^0_3} <Z_1 \otimes Z_2,\Delta(x_{i_{\tau_1}}x_{i_{\tau_2}}><Z_3,x_{i_3}>\\
&=& \Sigma_{\tau \in \Pi^0_3} \Sigma_{\sigma \in \Pi_2} < \phi_1,x_{i_{\tau_{\sigma_1}}}>
< \phi_2,x_{i_{\tau_{\sigma_2}}}><\phi_3,x_{\tau_3}>\\
&=& \Sigma_{\sigma \in \Pi_3}< \phi_1,x_{i_{\tau_{\sigma_1}}}>< \phi_2,x_{i_{\tau_{\sigma_2}}}>
< \phi_3,x_{i_{\tau_{\sigma_3}}}>\\
&=&  \Sigma_{\sigma \in \Pi_3}<\phi_{\sigma_1},x_{i_1}><\phi_{\sigma_2},x_{i_2}><\phi_{\sigma_3},x_{i_3}>.
\end{eqnarray*}

 It should now be clear how things propagate recursively to yield 
a proof for general $n$. We assume that Table $T_n$ is as follows: 
its $3[2(n-1)+1]=3(2n-1)$ tensor products consist of:

-- the $4(n-1)$ indicated by \# vanishing under $(Z_1 \otimes \cdots \otimes Z_n) \circ \Delta^{(n-1)}$

-- the $n+1$ indicated by \$ vanishing under $(Z_1 \otimes \cdots \otimes Z_n) \circ \Delta^{(n-1)}$ 

-- the remaining $n$ with sum $S_n = \Sigma_{\tau \in \Pi^0_n} \lambda_{\tau_1}\lambda_{\tau_2}\ldots\lambda_{\tau_{(n-1)}}\otimes \lambda_{\tau_n}$, $\tau$ in the set $\Pi^0_n$ of cyclic permutations of $\{1,\ldots,n\}$, 
with (\ref{3.113})  holding for $\lambda_n = x_{i_n} \in \V$.

 From what precedes it is clear that these features propagate 
from $n$ to $n+1$. In particular we have:
\begin{eqnarray*}
S_{n+1} &=& S_n(\lambda_{n+1} \otimes {\mathbf 1}) + (\lambda_1 \ldots \lambda_n \otimes {\mathbf 1})({\mathbf 1} \otimes \lambda_{n+1})\\
&=& \Sigma_{\tau \in \Pi^0_n} \lambda_{\tau_1}\lambda_{\tau_2}\lambda_{\tau_{(n-1)}}\lambda_{n+1} \otimes \lambda_{\tau_n} + (\lambda_1 \ldots \lambda_n \otimes \lambda_{n+1})\\
&=& \Sigma_{\tau \in \Pi^0_{n+1}} \lambda_{\tau_1} \lambda_{\tau_2}\ldots
\lambda_{\tau_n} \otimes \lambda_{\tau_{(n+1)}}
\end{eqnarray*}
and
\begin{eqnarray*}
\lefteqn{<Z_1 * \cdots * Z_{n+1},x_{i_1}\ldots x_{i_{n+1}}>}\\
&=&
<Z_1 \otimes \cdots \otimes Z_{n+1} (\Delta \otimes \id)
\Delta^{(n-2)}(x_{i_1}\ldots x_{i_{n+1}})>\\
&=& \Sigma_{\tau \in \Pi^0_n} <Z_1 \otimes \cdots \otimes Z_n,
 \Delta^{(n-2)}(x_{i_{\tau_1}}\ldots x_{i_{\tau_n}})><Z_{n+1},x_{i_{n+1}})>\\
&=& \Sigma_{\tau \in \Pi^0_n}\Sigma_{\sigma \in \Pi_n} < \phi_1,x_{i_{\tau_{\sigma_1}}}> \cdots <\phi_n,x_{i_{\tau_{\sigma_n}}}><\phi_{n+1},x_{i_{(n+1)}}>\\
&=& \Sigma_{\sigma \in \Pi_{n+1}} <\phi_1,x_{i_{\sigma_1}}> \cdots <\phi_1,x_{i_{\sigma_n}}>\\
&=& \Sigma_{\sigma \in \Pi_{n+1}} < \phi_{\sigma_1},x_{i_1}> \cdots < \phi_{\sigma_1},x_{i_n}>.
\end{eqnarray*}
\end{proof}

Our final result elucidates the structure of ${\H}$ by displaying 
it as in strict Hopf-algebra-duality with the enveloping (Hopf) 
algebra ${\U}({\L})$ of the Lie algebra ${\L}$, the latter isomorphic 
to the Hopf algebra ${\H}_{*}$ by the Milnor-Moore theorem.

\begin{theorem}\label{t3.5}
 Let $\H$ be a CMK Hopf 
algebra.  $\H$ is separated by ${\H}_{*}$: for $h{\in}{\H}$, 
$<\xi,h> = 0$ for all $\xi \in \H_*$ entails $h=0$.
\end{theorem}

\begin{proof}
Each $h{\in}{\H}$ is of the form $h=\lambda{\mathbf 1} + h^{(1)} + h^{(2)} + \cdots + h^{(n)}$
with $h^{(k)}{\in}{\V}^{\vee k}$, $k=1,\ldots,n$, $n{\in}\N$. 
We assume that $h$ vanishes under all elements ${\xi}{\in}{\H}_{*}$ 
and show that $h=0$: indeed, let $Z_i \leftrightarrow \phi_i \in \L$, $i=1,\ldots,n$: then:

-- since by (\ref{3.8}) all the terms of h but the first lie in ${\H}^{+}= $ker${\varepsilon}$ one 
has $<\{\mathbf 1_*,h> = \varepsilon(h) = \lambda$: the 
requirement  $<\{\mathbf 1_*,h> = 0$ thus entails the vanishing 
of the first term $\lambda {\mathbf 1}$.

-- next by Corollary \ref{c3.4}(ii)  one has $<Z_i,h> = \phi_i(h^{(1)})$: 
asking this to vanish for all $Z_{i}{\in}{\H}_{*}$ thus entails 
the vanishing of the second term $h^{(1)}$. 

-- assume that we have shown that $h^{(k)}$ vanishes for $k{\leq}p<n$ 
we have, by Corollary \ref{c3.4}(ii) and (iii): 
\begin{eqnarray*}
<Z_1 * \cdots Z_{p+1},h> &=& <Z_1 * \cdots Z_{p+1},h^{(p+1)} + \cdots + h^{(n)}>\\
&=& <Z_1 * \cdots Z_{p+1},h^{(p+1)}>\\
&=& (\phi_1 \vee \cdots \vee \phi_{p+1})(h^{(p+1)})
\end{eqnarray*}
whose vanishing under all $\phi_1 \vee \cdots \vee \phi_{p+1}$ entails 
$h^{(p+1)}=0$: we thus proved Theorem \ref{t3.5} inductively. 
\end{proof}

\section{Loops of CMK Hopf algebras}\label{sec4}

 In what follows ${\H}(m,{\mathbf 1} =e1,{\Delta},{\varepsilon},S)$ is a 
CMK Hopf algebra, cf. Sections \ref{sec1} and \ref{sec3} of which 
we adopt the notation.

\begin{definition}\label{def4.1}
 In what follows the ground 
field is $\C$ and we will handle algebras with and without unit. 
Since our general practice was to use the word: algebra to mean: 
unital algebra, we shall write ``algebra" (quote-unquote) 
to mean: not necessarily unital algebra over $\C$.

(i)\ We consider the following ``algebras" of analytic 
functions on the Riemann sphere {\bf PC}$^{1}$  with north pole $N={\infty}$ 
and south pole $S=0$.\footnote{Notation using the chart of {\bf PC}$^{1}$ 
obtained by stereographic projection from the north pole on the equator 
coordinatized by $z$: in phrasing the Definitions (\ref{4.3}) below, 
this chart is used to identify 
{\bf PC}$^{1}$ -- $\{0,{\infty}\}$ with $\C-\{0\}$.} 
\begin{eqnarray}\label{4.1}
\left\{\begin{array}{l}
\bA = \{f \in {\bf Holom}(\C-\{0\}) \mbox{ with 0 a pole of finite order} \\
\bA_- = \{\mbox{polynomials in $\frac{1}{z}$ without constant term}\}\\
\bA_+ = \{\mbox{Restrictions to $(\C-\{0\})$ of functions in {\bf Holom}$(\C)$}\}\\
\bA^\infty_- = \{\mbox{Series in $\frac{1}{z}$ without constant term}\}
\end{array}\right. .
\end{eqnarray}

(ii)\ A {\bf loop} is a map ${\gamma}: $ {\bf PC}$^{1}-\{0,{\infty}\}{\rightarrow}{\G}= Char\H$. 
The set of loops is denoted by $L({\H})$.

(iii)\ With ${\B}$ one of the algebras ${\bA}$, ${\bA}_{-}$, ${\bA}_{+}$, or 
${\bA}^\infty_-$, a ${\B}$-{\bf loop} is a map ${\gamma}: $ 
{\bf PC}$^{1}-\{0,{\infty}\}{\rightarrow}{\G}$ 
such that the functions 
{\bf PC}$^{1}-\{0,{\infty}\} {\ni}z{\rightarrow}{<}{\gamma}(z),h>$, $h{\in}{\H}^{+}= $ Ker${\varepsilon}$, belongs 
to ${\B}$.\footnote{Definition implying that $<{\gamma}(z),{\mathbf 1}>=1$ 
for all $z$.}
The set of ${\B}$-loops is denoted by $L_{\B}(\H)$.
\end{definition}

We recall that a ${\bA}$-loop ${\gamma}$ has a unique {\bf Birkhoff decomposition}:\\
\begin{eqnarray}\label{4.2}
\gamma = \gamma_-^{-1} \gamma_+,
\end{eqnarray}
where
\begin{eqnarray}\label{4.3}
\left\{\begin{array}{ll}
\C-\{0\} \ni z \to <\gamma_-(z),h> & \mbox{belongs to } \bA_-\\
\C-\{0\} \ni z \to <\gamma_+(z),h> & \mbox{belongs to } \bA_+
\end{array}\right. ,\ h \in \H^+.
\end{eqnarray}

\begin{lemma}\label{l4.2}
(i)\ The ${\B}$-loops build a group under the operations:
\begin{eqnarray}\label{4.4}
\left\{\begin{array}{l}
\mbox{product } \gamma'\gamma^{\prime\prime}: (\gamma'\gamma^{\prime\prime})(z) = \gamma'(z) * \gamma^{\prime\prime}(z),\ z \in {\bf PC}^1-\{\infty\}\\
\mbox{inverse }: \gamma^{-1}(z) = \gamma(z)^{-1} = S_*\gamma(z)
\end{array}\right.
\end{eqnarray}
with the unit the constant loop:  $z{\rightarrow}{<}{\mathbf 1}_{*},\cdot{>}={\varepsilon}$.

 (ii)\ The ${\B}$-loops $\gamma$ are one-to-one 
with elements $\ugamma{\in} $Hom$_{\mbox{``alg"}}({\H},{\B}')$ through 
the bijection:\footnote{Hom$_{\mbox{``alg"}}({\H},{\B}')$ is the set of homomorphisms between ``algebras", 
i.e. $\C$-linear multiplicative maps: ${\H}{\rightarrow}{\B}$.}
\begin{eqnarray}\label{4.5}
\gamma \leftrightarrow \ugamma: <\gamma(z),h> = \ugamma(h)(z), \quad h \in \H,\ z \in \C-\{0\},
\end{eqnarray}
a group isomorphism with the following correspondence of 
group products:
\begin{eqnarray}\label{4.6}
\gamma'\gamma^{\prime\prime} \leftrightarrow \ugamma = \ugamma' * \ugamma^{\prime\prime},
\end{eqnarray}
where the product on the left is that of the group ${\G}$ (first line 
(\ref{1.1a})), whilst $*$ on the right is the convolution 
product of Hom$_{\mbox{``alg"}}({\H},{\B}')$:
\begin{eqnarray}\label{4.7}
\left\{\begin{array}{l}
(\ugamma' * \ugamma^{\prime\prime})(h) = \ugamma'(h_{(1)})\ugamma^{\prime\prime}(h_{(2)})\\
(\ugamma)^{-1}(h) = \ugamma^{-1}(h) = \ugamma(Sh)
\end{array}\right. .
\end{eqnarray}
\end{lemma}

\begin{proof}
 One has for $h,h'{\in}{\H}$:
\begin{eqnarray*}
<\gamma(z),hh'> &=& <\gamma(z),h> \cdot <\gamma(z),h'> = \ugamma(h)(z)\ugamma(h')(z)\\
&=& [\ugamma(h)\ugamma(h')] (z) = \ugamma(hh')(z),\\
<\gamma(z),{\mathbf 1}> &=& 1 = \ugamma({\mathbf 1})(z) = {\mathbf 1}_{\B} = 1,\\
(\ugamma' * \ugamma^{\prime\prime})(h)(z) &=&
<(\gamma'\gamma^{\prime\prime})(z),h>\\
&=& <\gamma'(z)*\gamma^{\prime\prime}(z),h> = <\gamma'(z) \otimes \gamma^{\prime\prime}(z),\Delta h>\\
&=& <\gamma'(z) \otimes \gamma^{\prime\prime}(z),h_{(1)} \otimes h_{(2)}> = <\gamma'(z),h_{(1)}><\gamma^{\prime\prime}(z),h_{(2)}>\\
&=& \ugamma'(h_{(1)})(z)\ugamma^{\prime\prime}(h_{(2)})(z) = [\ugamma'(h_{(1)})\ugamma^{\prime\prime}(h_{(2)})](z).
\end{eqnarray*}
\end{proof}

In the sequel we are concerned with the ``special loops" defined as follows: 

\begin{definition}\label{def4.3}
 The ${\B}$-loop ${\gamma}$, 
${\B}= {\mathbf A}_{-}$ or ${\mathbf A}^{\infty}_{-}$, is {\bf special} whenever its inverse ${\phi}$:\footnote{ 
To alleviate writing we write the complex variable of the loop 
${\phi}$ as an index. Note that the limit in (\ref{4.9}) is taken 
in the topology of ${\G}$: simple convergence on ${\H}$.}
\begin{eqnarray}\label{4.8}
\phi_\varepsilon = \gamma{(\varepsilon)^*}^{-1} \quad \mbox{(inverse in the group $\G$)}
\end{eqnarray}
is such that ${\phi_\varepsilon^*}^{-1} *\theta_{t\varepsilon}(\phi_\varepsilon)$,
(product in the group ${\G},t,{\varepsilon}{\in}\R)$ has a limit 
for ${\varepsilon}{\rightarrow}0$ for all  $t$:
\begin{eqnarray}\label{4.9}
{\phi_\varepsilon^*}^{-1} * \theta_{t\varepsilon}(\phi_\varepsilon) \underset{\varepsilon=0}{\rightarrow} F_t,
\end{eqnarray} 
and moreover $F_{t}$ is differentiable in $t$ for $t=0$.
\end{definition}

\begin{remark}\label{rem4.4}
 In the case of ${\mathbf A}_{-}$-loops, as 
soon as condition (\ref{4.9}) is fulfilled, $F_{t}$ is differentiable as 
shown by the following computation. Continuity of the limit needs 
only be examined for the values on the kernel of ${\varepsilon}$ since 
$<{\phi_\varepsilon^*}^{-1} * \theta_{t\varepsilon}(\phi_\varepsilon),{\mathbf 1}> = 1$ 
is constant. Now for $h{\in}{\H}^{+}$ with ${\delta}_{Y}(h)=p$:
\begin{eqnarray*}
\lefteqn{
<{\phi_\varepsilon^*}^{-1} * \theta_{t\varepsilon}(\phi_\varepsilon),h>}\\
&=& <{\phi_\varepsilon^*}^{-1} \otimes \theta_{t\varepsilon}(\phi_\varepsilon),\Delta h>\\
&=& <\phi_\varepsilon \otimes \phi_\varepsilon,(S \otimes \theta_{t\varepsilon})({\mathbf 1} \otimes h + h \otimes {\mathbf 1} + h' \otimes h^{\prime\prime})>\\
&=& <\phi_\varepsilon \otimes \phi_\varepsilon,{\mathbf 1} \otimes \theta_{t\varepsilon}(h) + Sh\otimes {\mathbf 1} + Sh'\otimes\theta_{t\varepsilon}h^{\prime\prime}>\\
&=& <\phi_\varepsilon \otimes \phi_\varepsilon,{\mathbf 1} \otimes e^{tp\varepsilon}h + Sh\otimes {\mathbf 1} + Sh'\otimes e^{tp^{\prime\prime}\varepsilon}h^{\prime\prime}>\\
&=&<\phi_\varepsilon,{\mathbf 1}>e^{tp\varepsilon}h<\phi_\varepsilon,h> + <\phi_\varepsilon,Sh><\phi_\varepsilon,{\mathbf 1}>\\
&&  + <\phi_\varepsilon,Sh'>e^{tp^{\prime\prime}\varepsilon}<\phi_{\varepsilon},h^{\prime\prime}>\\
&=& e_{tp\varepsilon}P\left(\frac{1}{\varepsilon}\right) + P_1\left(\frac{1}{\varepsilon}\right)
+ e^{tp^{\prime\prime}\varepsilon}P'\left(\frac{1}{\varepsilon}\right) P^{\prime\prime}\left(\frac{1}{\varepsilon}\right).
\end{eqnarray*}
This makes it clear that the limit $F_{t}$ is a polynomial in $t$ if 
it exists. 
\end{remark}

\begin{lemma}\label{l4.5}
Let $\gamma$ be a special 
loop as in  Definition \ref{def4.3}. Then \\
\ $F_{t}$ in (\ref{4.9}) is a flow, one has:
\begin{eqnarray}\label{4.10}
F_{t+s} = F_t * F_s,
\end{eqnarray}
thus
\begin{eqnarray}\label{4.11}
F_t = e^{*\beta t},
\end{eqnarray}
with the ${\beta}$-{\bf function}\footnote{terminology borrowed 
from renormalization theory.} the following element of $\L$:
\begin{eqnarray}\label{4.11a}
\beta = \frac{\partial e^{*\beta t}}{\partial t} \biggr|_{t=0}.
\end{eqnarray}
\end{lemma}

\begin{proof}
(i)\ Let $t,s{\in}\R$. First $$\theta_{* t \varepsilon}[{\phi_\varepsilon^*}^{-1}\theta_{*s\varepsilon}(\phi_\varepsilon)]\underset{\varepsilon=0}{\rightarrow} F_s$$
indeed, for $h{\in}{\H}$ of degree $n$:
first 
\begin{eqnarray*}
<\theta_{* t \varepsilon}[{\phi_\varepsilon^*}^{-1}*\theta_{*s\varepsilon}(\phi_\varepsilon)],h>
&=& <[{\phi_\varepsilon^*}^{-1}*\theta_{*s\varepsilon}(\phi_\varepsilon)],\theta_{* t \varepsilon}h>\\
&=& e^{int \varepsilon} <{\phi_\varepsilon^*}^{-1} *\theta_{*s\varepsilon}(\phi_\varepsilon),h> \rightarrow F_s
\end{eqnarray*}
since $e^{int \varepsilon} \rightarrow 1$. Then:
\begin{eqnarray*}
F_{t+s} &=& \lim_{\varepsilon=0}
{[}{\phi_\varepsilon^*}^{-1}*\theta_{*(t+s)\varepsilon}(\phi_\varepsilon)]\\
&=& \lim_{\varepsilon=0} [{\phi_\varepsilon^*}^{-1}*\theta_{*t\varepsilon}[\theta_{*s\varepsilon}(\phi_\varepsilon)]]\\
&=& \lim_{\varepsilon=0}
[{\phi_\varepsilon^*}^{-1}*\theta_{*t\varepsilon}\{\phi_\varepsilon * {\phi_\varepsilon^*}^{-1}* \theta_{*s\varepsilon}(\phi_\varepsilon)\}]\\
&=& \lim_{\varepsilon=0}
[{\phi_\varepsilon^*}^{-1}*\theta_{*t\varepsilon}\phi_\varepsilon * \theta_{*t\varepsilon}[{\phi_\varepsilon^*}^{-1}* \theta_{*s\varepsilon}(\phi_\varepsilon)]]\\
&=& \{\lim_{\varepsilon=0}
{\phi_\varepsilon^*}^{-1}*\theta_{*t\varepsilon}\phi_\varepsilon\} * \{\lim_{\varepsilon=0} \theta_{*t\varepsilon}[{\phi_\varepsilon^*}^{-1} \theta_{*s\varepsilon}(\phi_\varepsilon)]\\
&=& F_t * F_s.
\end{eqnarray*}
\end{proof}

\begin{lemma}\label{l4.6}
Let ${\gamma}$ be a special $\bA^\infty_-$-loop 
with:
\begin{eqnarray}\label{4.12}
\phi_\varepsilon = {\mathbf 1}_* + \Sigma_{n \geq 1} \frac{d_n}{\varepsilon^n} \quad \mbox{with } d_n \in \H^{*+}, \mbox{ i.e., } <d_n,{\mathbf 1}> = 0,\ n \geq 1
\end{eqnarray}
(we set $d_{1}= $ Res${\phi}$, called the {\bf residue of} ${\phi}$).

One has:
\begin{eqnarray}\label{4.13}
\left\{\begin{array}{l}
Y_*d_1 = Y_*(\mbox{Res}\phi) = \beta\\
Y_*d_{n+1} = d_n * \beta, \ n > 1
\end{array}\right. ,
\end{eqnarray}
rewritten
\begin{eqnarray}\label{4.13a}
&& \left\{\begin{array}{l}
d_1 = {\underline{Y}_*}^{-1}\beta\\
d_{n+1} = {\underline{Y}_*}^{-1} d_n * \beta,\ n > 1
\end{array}\right. ,\nonumber\\
&& \qquad\qquad  {\underline{Y}_*} \mbox{ the injective restriction of $Y_*$ to } \H^{*+} = \mbox{ Ker}\varepsilon_*,
\end{eqnarray}
explicitly:
\begin{eqnarray}\label{4.13b}
d_n = \int_{s_1 \geq s_2 \geq \cdots \geq s_n \geq 0}
\theta_{*-s_1}(\beta) * \theta_{*-s_2}(\beta) * \cdots * \theta_{*-s_n}(\beta)\ ds_1 \ldots ds_n
\end{eqnarray}
(using
\begin{eqnarray}\label{4.14}
{\underline{Y}_*}^{-1} = \int^\infty_0 \theta_{*-s}\ ds \mbox{ in restriction to } \H^{*+}).
\end{eqnarray}
\end{lemma}

The morale of this is that the loop ${\phi}$ is determined by 
its residue, thus by ${\beta}$, with dependence expressed explicitly 
by (\ref{4.13b}).

\begin{proof}
Performing a licit exchange of limits in:
\begin{eqnarray}\label{4.15a}
\begin{array}{ccc}
<{\phi_\varepsilon^*}^{-1} *(\theta_{t\varepsilon}\phi_\varepsilon),h> & \overset{{\displaystyle\lim_{\varepsilon=0}}}{\rightarrow} & <e^{\beta t},h>\\[2mm]
\downarrow \frac{\partial}{\partial t} \big|_{t=0} &&
\downarrow \frac{\partial}{\partial t} \big|_{t=0}\\[1mm]
\varepsilon<{\phi_\varepsilon^*}^{-1} *(Y\phi_\varepsilon),h> & \overset{{\displaystyle\lim_{\varepsilon=0}}}{\rightarrow} & <\beta,h>
\end{array} ,\ h \in \H,
\end{eqnarray}
we first show that we have
\begin{eqnarray}\label{4.15b}
<\beta,h> = \lim_{\varepsilon=0} \varepsilon<{\phi_\varepsilon^*}^{-1} *
  Y_*(\phi_\varepsilon),h> , \qquad h \in \H,
\end{eqnarray}
Indeed, the horizontal limit comes from (\ref{4.9}) with $F_{t}=e^{*{\beta}t}$; 
the right vertical limit is (\ref{4.11a}). As for the left vertical 
limit, we have that  $\frac{\partial}{\partial t}|_{t=0}$ applied to: 
\begin{eqnarray*}
<{\phi_\varepsilon^*}^{-1} *(\theta_{t\varepsilon}\phi_\varepsilon),h> &=&
<(S_*\phi_\varepsilon)\otimes(\theta_{* t \varepsilon}\phi_\varepsilon),h>\\
&=& <\phi_\varepsilon \otimes \phi_\varepsilon,(S \otimes e^{*Y_*\varepsilon t}\phi_\varepsilon),\Delta h>\\
&=& <\phi_\varepsilon,Sh_{(1)}><\phi_\varepsilon,e^{Yt}h_{(2)}>
\end{eqnarray*}
yields
\begin{eqnarray*}
<\phi_\varepsilon, Sh_{(1)}><\phi_\varepsilon,Yh_{(2)}> &=& \varepsilon<S_*\phi_\varepsilon,h_{(1)}><Y\phi_\varepsilon,Yh_{(2)}>\\
&=& e<{\phi_\varepsilon^*}^{-1} \otimes(Y_* \phi_\varepsilon),\Delta h>\\
&=& \varepsilon<{\phi_\varepsilon^*}^{-1}*(Y_*\phi_\varepsilon),h>.
\end{eqnarray*}

Next the function $\{\varepsilon \to \varepsilon<{\phi_\varepsilon^*}^{-1} * Y_*(\phi_\varepsilon),h>\}$ is 
holomorphic on the whole Riemann sphere, thus must be constant: 
we have   $\varepsilon<{\phi_\varepsilon^*}^{-1} * Y_*(\phi_\varepsilon),h> = <\beta,h>$, ${\phi_\varepsilon^*}^{-1} * Y_*(\phi_\varepsilon) = \frac{1}{\varepsilon} \beta$,
i.e.:
\begin{eqnarray}\label{4.16}
Y_*\phi_\varepsilon = \phi_\varepsilon * \frac{1}{\varepsilon} \beta.
\end{eqnarray}
Feeding (\ref{4.12}) into this yields: 
\begin{eqnarray*}
\sum_{n \geq 1} \frac{1}{\varepsilon^n} Y_*d_n &=& \frac{1}{\varepsilon} {\mathbf 1} * \beta + \sum_{n \geq 1} \frac{d_n}{\varepsilon^{n+1}} * \beta\\
&=& \frac{1}{\varepsilon} \beta + \sum_{n \geq 2} \frac{d_{n-1}}{\varepsilon^n} * \beta,
\end{eqnarray*}
 equating 
coefficients of $\frac{1}{\varepsilon^n}$ then yields (\ref{4.13}).

Passage from (\ref{4.13}) to (\ref{4.13a}): $Y_{*}$ a is injective in restriction 
to the augmentation ideal ${\H}^{*+}={\oplus}_{n{\geq}1}{\H}^{*n}$. 
Thus ${\beta}$ (thus also Res${\phi})$ determines $F$, the key 
to the explicit dependence being formula (\ref{4.14}), itself obtained 
as follows: integrating from 0 to $\infty$,
\begin{eqnarray*}
\frac{\partial \theta_{*-s}}{\partial t} = \frac{\partial e^{*-tY_*}}{\partial t}
 = -Y_*e^{*-tY_*} = -Y_*\theta_{*-s}
\end{eqnarray*} 
yields $-\id_{\H^*}=Y_{*}\int^\infty_0 {\theta}_{*-s}\, ds$,
 it suffices now 
to inverse $Y_{*}$ in restriction to ${\H}^{*+}$. Applying (\ref{4.14}) to 
${\beta}$ yields (\ref{4.13b}) for $n=1$. Inductive iteration then yields 
(\ref{4.13b}): assume it holds for $n$: applying ${\underline{Y}}^{-1}$ on both 
sides yields: 
\begin{eqnarray*}
d_{n+1} &=& {\underline{Y}}^{-1}d_n * \beta \\
&=& \int^\infty_0 \theta_{*-s} \biggl[\int_{s_1\geq s_2 \geq \cdots \geq s_n \geq 0}
\theta_{*-s_1}(\beta) * \theta_{*-s_2}(\beta) \\
&& \qquad * \cdots * \theta_{*-s_n}(\beta)\ ds_1 \ldots ds_n\biggr]ds\\
&=& \int^\infty_0  \biggl[\int_{s_1\geq s_2 \geq \cdots \geq s_n \geq 0}
\theta_{*-(s+s_1)}(\beta) * \theta_{*-(s+s_2)}(\beta) \\
&& \qquad * \cdots * \theta_{*-(s+s_n)}(\beta)\ ds_1 \ldots ds_n\biggr]ds\\
&=& \int_{s\geq s_1\geq s_2 \geq \cdots \geq s_n \geq 0}
\theta_{*-s}(\beta) * \theta_{*-s_1}(\beta) *\theta_{*-s_2}(\beta) \\
&& \qquad * \cdots * \theta_{*-s_n}(\beta)\ dsds_1 \ldots ds_n
\end{eqnarray*}
up to numbering of variables identical to (\ref{4.13b}).
\end{proof}

Extension of the Lie algebra ${\L}$ by an additional element $Z_{0}$
enables 
us now to give a compact expression of the last result, and, 
as a byproduct, a characterization of ``specialty" for ${\mathbf A}^\infty_-$-loops.

\begin{def-lemma}\label{def4.7}
 (i)\  $\tilde{\L}$ is ${\L}{\oplus}\C\{Z_{0}\}$ with the Lie-brackets of ${\L}$ plus:
\begin{eqnarray}\label{4.17}
{[}Z_0,Z] = -[Z,Z_0] = Y_*(Z), \qquad Z \in \L.
\end{eqnarray}

(ii)\ The so defined $\tilde{\L}$ is a Lie algebra.
\end{def-lemma}

\begin{proof} We need only check the additional cases of the 
Jacobi requirement. Now, for $Z,Z'{\in}{\L}$, we have $$[Z_{0},[Z_{0},Z]]+[Z_{0},[Z,Z_{0}]]+[Z,[Z_{0},Z_{0}]]=0,$$ 
and  
$$[Z_{0},[Z,Z']]+[Z,[Z',Z_{0}]]+[Z'[Z_{0},Z]]=Y_{*}([Z,Z'])-[Z,Y_{*}(Z')]+[Z',Y_{*}(Z)]=0$$
by the fact that the algebra-derivation $Y_{*}$ of ${\H}^*$ is a Lie 
algebra-derivation of the Lie algebra Lie${\H}^*$: for ${\xi},{\eta}{\in}{\H}^*$ 
we have
$Y_*(\xi\eta) = (Y_*\xi)\eta + \xi(\eta Y_*)$ and $Y_*(\eta\xi) = (Y_*\eta)\xi + \eta(\xi Y_*)$
thus 
$Y_*[\xi,\eta] = [Y_*\xi,\eta] + [\xi,\eta Y_*]$.
\end{proof}

\begin{theorem}\label{t4.8}
With $\tilde{\L}$ as in Definition-Lemma \ref{def4.7}, one 
has for each ${\mathbf A}^\infty_-${-loop} ${\phi}$
\begin{eqnarray}\label{4.18}
\phi_\varepsilon = \lim_{t = \infty} e^{*-tZ_0} * e^{*-t(\beta/\varepsilon+Z_0)} \qquad \mbox{where }
\beta = Y(\mbox{Res}\phi).
\end{eqnarray}
\end{theorem}

\begin{proof}
 We apply the expansional formula
\begin{eqnarray}\label{4.19}
e^{*A+B} = \sum_{n \in \N} \int_{\Sigma u_i = 1,u_i>0} e^{*u_0A}*B*e^{*u_1A} * \cdots * B * e^{*u_nA}du_1\ldots du_n
\end{eqnarray}
to $A = tZ_0$, $B = t\beta$, $t > 0$, yielding:
\begin{eqnarray*}
\lefteqn{e^{*t(\beta+Z_0)}}\\
 &=& \sum_{n\in \N} \int_{\Sigma u_i = 1,u_i > 0} e^{*u_0tZ_0} * \beta * 
    e^{*u_1tZ_0} * \cdots  \\
 & &  \qquad\qquad\qquad\qquad *e^{*u_{n-1}tZ_0}*\beta * e^{*u_ntZ_0}t^n  du_1 \ldots du_n \\
 & & \qquad(v_i = tu_i, \ i = 1,\ldots,n)\\
&=& \sum_{n \in \N} \int_{\Sigma v_i=t,v_i > 0} e^{*v_0Z_0} * \beta * e^{*v_1Z_0} * \cdots * \beta * e^{*v_nZ_0}dv_0dv_1\ldots dv_n\\
&& \qquad  (v_0=t-s_1,\ v_1 = s_1-s_2,\ldots,v_{n-1}=s_{n-1}-s_n,\ v_n = s_n,\ \beta \to \frac{1}{\varepsilon}\beta)
\end{eqnarray*}
\begin{eqnarray*}
\lefteqn{
e^{*-t(\beta/\varepsilon+Z_0)}}\\
&=& \frac{1}{\varepsilon^n} \sum_{n \in \N} \int_{t \geq s_1 \geq s_2 \geq \cdots \geq s_n \geq 0}
e^{*(t-s_1)Z_0} * \beta * e^{*(s_1-s_2)Z_0} * \cdots \\
&& \qquad\qquad\qquad * e^{*(s_{n-1}-s_n)} * \beta * e^{*s_nZ_0} dv_0dv_1\ldots dv_n\\
&=& e^{*tZ_0} \sum_{n \in\N} \frac{1}{\varepsilon^n} \int_{t \geq s_1 \geq s_2 \geq \ldots \geq s_n \geq 0}
\theta_{*-s_1}(\beta)* \theta_{*-s_2}(\beta) * \cdots \\
&& \qquad\qquad\qquad * \theta_{*-s_n}(\beta) dv_0dv_1\ldots,dv_n
\end{eqnarray*} 

We thus have
\begin{eqnarray*}
\lefteqn{
e^{*-tZ_0} * e^{*-t(\beta/\varepsilon + Z_0)}}\\
&=& \sum_{n \in \N} \frac{1}{\varepsilon^n} \int_{t\geq s_1 \geq s_2 \geq \cdots \geq s_n \geq 0}
\theta_{*-s_1}(\beta)* \theta_{*-s_1}(\beta) * \cdots \\
&&\qquad\qquad\qquad * \theta_{*-s_n}(\beta) dv_0dv_1\ldots,dv_n,
\end{eqnarray*}
whence (\ref{4.18})
\begin{eqnarray*}
\lefteqn{\lim_{t=\infty} e^{*-tZ_0} * e^{*-t(\beta/\varepsilon+Z_0)}}\\
&=& \sum_{n \in \N} \frac{1}{\varepsilon^n}\int_{s_1 \geq s_2 \geq \cdots \geq s_n \geq 0}
\theta_{*-s_1}(\beta)* \theta_{*-s_1}(\beta) * \cdots * \theta_{*-s_n}(\beta) dv_0dv_1\ldots,dv_n\\
&=& \sum_{n \in \N} \frac{1}{\varepsilon^n} d_n = \phi_\varepsilon.
\end{eqnarray*}
\end{proof}

\section{Algebra\"{\i}c Birkhoff decomposition}\label{sec5}

\begin{definition}\label{def5.1}
 A {\bf Birkhoff sum} (over ${\k}$) is 
a commutative (unital) ${\k}$-algebra ${\A}$ which is the direct 
sum ${\A}={\A}_{-}{\oplus}{\A}_{+}$ of two ${\k}$-linear multiplicative 
subspaces ${\A}_{-}$ and ${\A}_{+}$.\footnote{We do not say: subalgebras 
because according to our general terminological practice a subalgebra 
contains the unit. Note that Definition (\ref{def5.1}) follows from the 
fact that both ${\A}_{-}$ and ${A}_{+}$ are closed under multiplications.}
 The projection $T: {\A}{\rightarrow}{\A}_{-}$ parallel to ${\A}_{+}$ thus 
fulfills:
\begin{eqnarray}\label{5.1}
T(ab) + (Ta)(Tb) = T[(Ta)b + a(Tb)], \qquad a,b \in \A.
\end{eqnarray}
\end{definition}

\begin{proposition}\label{p5.2}
 {\em (Algebra\"{\i}c Birkhoff decomposition).} 
With $\H$ a $\N$-graded, connected, progressive,\footnote{i.e., ${\H}$ has a $\N$-grading commuting with the Hopf structure, such that ${\H}_{0}=\C{\mathbf 1}$, and moreover such that
$\Delta h = h \otimes {\mathbf 1} + {\mathbf 1} \otimes h + h' \otimes h^{\prime\prime}$
with max(degh$'$,degh$^{\prime\prime}) < p$, $h \in H^p,\ p > 0$.}
commutative Hopf algebra over ${\k}$, and ${\A}={\A}_{-}{\oplus}{\A}_{+}$ a 
Birkhoff sum, denote by Hom$_{\k\mbox{-alg}}({\H},{\A}$) the 
set of unital ${\k}$-algebra homomorphisms: ${\H}{\rightarrow}{\A}$, 
considered as a ${\k}$-algebra under the convolution product:\footnote{The multiplicativity of ${\phi}$ and ${\phi}'$ implies that 
of ${\phi}{\phi}'$ owing to the commutativity of ${\A}$.}
\begin{eqnarray}\label{5.2}
(\phi\phi')(h) = m_{\A}(\phi \otimes \phi')(\Delta h), \quad
\phi,\phi' \in \mbox{ Hom}_{\k\mbox{-alg}}(\H,\A),\ h \in H.
\end{eqnarray}

Let $\phi \in \mbox{ Hom}_{\k\mbox{-alg}}(\H,\A)$ be given.  Requiring, for a $\k$-linear $\phi_-:\H\to \A$:
\begin{eqnarray}\label{5.3a}
\phi_-({\mathbf 1}) = {\mathbf 1}_{\A}
\end{eqnarray}
\begin{eqnarray}\label{5.3b}
\phi_-(h) = -T[\phi(h) + \phi_-(h')\phi(h^{\prime\prime})],
\end{eqnarray}
$h \in \H^+ = $ Ker$\varepsilon$,\footnote{The (commutative) 
product ${\phi}_{-}(h'){\phi}(h^{\prime\prime})$ r.h.s.\ is in ${\A}$.} 
specifies inductively a $\phi_- \in \mbox{ Hom}_{\k\mbox{-alg}}(\H,\A)$
having its range in ${\A}_{-}$:
\begin{eqnarray}\label{5.4}
\phi_-(h) = T\phi_-(h),
\end{eqnarray}
such that the product $\phi_+ = \phi_-\phi \in \mbox{ Hom}_{\k\mbox{-alg}}(\H,\A)$ has its range in ${\A}_{+}$:
\begin{eqnarray}\label{5.5}
\phi_+(h) = (\id_{\A} - T)[\phi(h) + \phi_-(h')\phi(h^{\prime\prime})], \qquad h \in \H^p,
\end{eqnarray}
this yielding the ``algebra\"{\i}c Birkhoff decomposition":\footnote{The product and the inverse r.h.s. are taken w.r.t. the convolution 
product (\ref{5.2}).}
\begin{eqnarray}\label{5.6}
\phi = {\phi_-}^{-1}\phi_+.
\end{eqnarray}
\end{proposition}

\begin{proof}
 Progressiveness of ${\H}$ implies that (\ref{5.3a}) and (\ref{5.3b}) 
specifies inductively a ${\k}$-linear map ${\phi}_{-}$: ${\H}{\rightarrow}{\A}$.

Check of (\ref{5.4}): obvious from (\ref{5.3b}) and $T^{2}=T$.

Check of (\ref{5.5}): we have, by (\ref{5.2}) and (\ref{5.3a}),(\ref{5.3b}), for $h{\in}{\H}$, 
omitting parentheses:
\begin{eqnarray*}
(\phi_-\phi')(h) &=& m_{\A}(\phi_- \otimes \phi)(\Delta h) \\
&=& m_{\A}(\phi_- \otimes \phi)(h \otimes {\mathbf 1} + {\mathbf 1} \otimes h + h' \otimes h^{\prime\prime})\\
&=& \phi_-h + \phi h + \phi_-h' \cdot \phi h^{\prime\prime},
\end{eqnarray*}
$$T[(\phi_-\phi')(h)] = \phi_-h + T[\phi_-h + \phi h + \phi_-h' \cdot \phi h^{\prime\prime}] = 0,$$
whence (\ref{5.5}) owing to $T^{2}=T$.

What is not obvious from (\ref{5.3b}) is that ${\phi}_{-}$ is multiplicative. 
We prove this by induction: for $x{\in}{\H}^{p}$, $y{\in}{\H}^{p}$ we 
have, still omitting parentheses:
\begin{eqnarray*}
\begin{array}{lclll}
\Delta x &=& x \otimes {\mathbf 1} + {\mathbf 1} \otimes x + x' \otimes x^{\prime\prime} &\mbox{with} & \deg x',\ \deg x^{\prime\prime} < p\\
\Delta y &=& y \otimes {\mathbf 1} + {\mathbf 1} \otimes y + y' \otimes y^{\prime\prime} & \mbox{with} & \deg y', \deg y^{\prime\prime} < q\\
\phi_-x &=& -T[\phi x + \phi_-x' \cdot \phi x^{\prime\prime}] = -TA & \mbox{with} & A = \phi x + \phi_-x' \cdot \phi x^{\prime\prime}\\
\phi_-y &=& -T[\phi y + \phi_-y' \cdot \phi y^{\prime\prime}] = -TB & \mbox{with} & B = \phi y + \phi_-y' \cdot \phi y^{\prime\prime}
\end{array}
\end{eqnarray*}
\begin{eqnarray*}
\Delta xy &=& xy \otimes {\mathbf 1} + {\mathbf 1} \otimes xy + x \otimes y + y \otimes x + x'y \otimes x^{\prime\prime} + x' \otimes x^{\prime\prime} y + xy'\otimes y^{\prime\prime}\\ 
&&+ y' \otimes xy^{\prime\prime}
 + x'y'\otimes x^{\prime\prime}y^{\prime\prime}
\end{eqnarray*}
hence, by the induction hypothesis:
\begin{eqnarray*}
\phi_-(xy) &=& -T[\phi xy + \phi_-x \cdot \phi y + \phi_-y \cdot \phi x + \phi_-x'y \cdot \phi x^{\prime\prime} + \phi_-x' \cdot \phi x^{\prime\prime} y\\
&& + \phi_-xy' \cdot \phi y^{\prime\prime} + \phi_-y' \cdot \phi xy^{\prime\prime} + \phi_-x'y' \cdot \phi x^{\prime\prime}y^{\prime\prime}]\\
&& \qquad \quad 1 \qquad\qquad  2 \qquad\qquad\quad 3 \qquad\qquad 4 \qquad\qquad\quad 5\\
&=& -T[\phi x \cdot \phi y\! +\! \phi_-x \cdot \phi y + \phi_-y \cdot \phi x +\! \phi_-y \cdot \phi_-x' \cdot \phi x^{\prime\prime}\! +\! \phi y \cdot \phi_-x' \cdot \phi x^{\prime\prime}\\
&& \qquad\quad 6  \qquad \qquad\quad\qquad 7 \qquad\qquad\qquad 8 \\
&& + \phi_-x \cdot \phi_-y' \phi y^{\prime\prime} + \phi x \cdot \phi_-y' \cdot \phi y^{\prime\prime} + \phi_-x' \cdot \phi_-y' \cdot \phi x^{\prime\prime} \cdot y^{\prime\prime}].
\end{eqnarray*}

On the other hand we have, using (\ref{5.1}) and the commutativity of 
${A}$:
\begin{eqnarray*}
\phi_-x \cdot \phi_-y &=& TA \cdot TB = T[ATB + BTA - AB] = -T[A\phi_- y + B \phi_-x  + AB]\\
&=& -T[(\phi x + \phi_-x' \cdot \phi x^{\prime\prime}) \phi_-y + (\phi y + \phi_-y' \cdot \phi y^{\prime\prime})\phi_-x\\
&& + (\phi x + \phi_-x' \cdot \phi x^{\prime\prime})(\phi y + \phi_- y' \cdot \phi y^{\prime\prime})]\\
&& \qquad\qquad 3 \qquad\qquad 4 \qquad\qquad\qquad\qquad 2 \qquad\qquad 6\\
&=& -T[\phi_- y \cdot\phi x + \phi_-y \cdot \phi_-x' \cdot \phi x^{\prime\prime} + \phi_-x \cdot \phi y + \phi_- x \cdot \phi_- y' \cdot \phi y^{\prime\prime}\\
&& \qquad\quad 1 \qquad\qquad 7 \qquad\qquad\qquad 5 \\
&& + \phi x \cdot \phi y + \phi x \cdot \phi_-y' \cdot \phi y^{\prime\prime} + \phi y \cdot \phi_-x' \cdot \phi x^{\prime\prime} \\
&& \qquad\qquad 8\\
&& + \phi_-x' \cdot \phi x^{\prime\prime} \cdot \phi x^{\prime\prime} \cdot \phi y^{\prime\prime}].
\end{eqnarray*} 
the two expressions containing the same terms.
\end{proof}

\appendix
\section{Algebras, coalgebras, bialgebras, Hopf algebras}

 In what follows ${\k}$ is a commutative field of characteristics 
zero (e.g. $\C$, $\R$, or $\Q$).
\begin{def-notation}\label{defA.0}
 (Algebras, coalgebras, 
bialgebras, Hopf algebras).
Let ${\H}$ be a ${\mathbf k}$-vector space: with the following notation 
for the maps $m,{\Delta}, e, {\varepsilon}, S$:

\medskip

\noindent
{\bf Notation:}
\begin{eqnarray*}
\begin{array}{ccccc}
m: \H \otimes \H \to \H & \Delta: \H \to \H \otimes \H & e: \k \to \H & \varepsilon: \H \to \k & S: \H \to \H\\
\mbox{multiplication} & \mbox{comultiplication} & \mbox{unit} & \mbox{counit} & \mbox{antipode}\\
& \Delta a = a_{(1)} \otimes a_{(2)} & e(1) = {\mathbf 1}
\end{array}
\end{eqnarray*}
the following axioms successively define ${\mathbf H}$ as an algebra, 
coalgebra, bialgebra, and Hopf algebra:

\medskip

\noindent{{\bf Axioms}:}
 
\medskip

\noindent{\bf Algebra}
\begin{eqnarray*}
\begin{array}{cccc}
(Am) & m(\id_{\H} \otimes m)=m(m \otimes \id_{\H}) & (\H \otimes \H \to \H) & a(bc) = (ab)c\\
&&&\\
(Ae) & \left\{\begin{array}{l}
m(\id_{\H} \otimes e) = \id_{\H}\\
m(e \otimes \id_{\H}) = \end{array}\right.
& \left(\left\{\begin{array}{l}
\H \otimes \k \cong \H \to \H\\
\k \otimes \H \cong \H \to \H \end{array}\right.\right) &
\left\{\begin{array}{l}
a({\mathbf 1}k) = ka\\
({\mathbf 1}k) a = 
\end{array}\right.
\end{array}
\end{eqnarray*}

\medskip

\noindent
{\bf Coalgebra}
\begin{eqnarray*}
\begin{array}{cccc}
(C\Delta) & \Delta^{(2)} = (\id_{\H} \otimes \Delta)\Delta & (\H \to \H \otimes \H) & a_{(1)} \otimes a_{(2)(3)}\otimes a_{(2)(4)}\\
& \qquad= (\Delta \otimes \id_{\H})\Delta && =\! a_{(1)(3)}\!\otimes\! a_{(1)(4)}\! \otimes \!a_{(2)}\\
&&&\\
(C\varepsilon) & \left\{\begin{array}{l}
(\id_{\H} \otimes \varepsilon)\Delta = \id_{\H}\\
(\varepsilon \otimes \id_{\H})\Delta =
\end{array}\right.
& \left(\left\{\begin{array}{l}
\H \to \H \otimes \k \cong \H\\
\H \to \k \otimes \H \cong \H
\end{array}\right.\right) &
\left\{\begin{array}{l}
a_{(1)}\varepsilon(a_{(2)}) = a\\
\varepsilon(a_{(1)})a_{(2)} = a
\end{array}\right.
\end{array}
\end{eqnarray*}

\medskip

\noindent
{\bf Bialgebra:} add to the previous axioms:
\begin{eqnarray*}
\begin{array}{cccc}
(Bm) & \Delta m = m_{\otimes} (\Delta \otimes \Delta) & (\H \otimes \H
\to \H \otimes \H) & \Delta(ab) = (\Delta a)(\Delta b)\\
&&&\\
(Be) & \Delta e = e \otimes e & (\k \cong \k \otimes \k \to \H \otimes \H) & \Delta(k {\mathbf 1} =  k{\mathbf 1} \otimes {\mathbf 1}\\
&&&\\
(B\varepsilon) & \varepsilon m = \varepsilon \otimes \varepsilon & (\H \otimes \H \to \k \cong \k \otimes \k) & \varepsilon(ab) = (\varepsilon a)(\varepsilon b)\\
&&&\\
(B\varepsilon e) & \varepsilon e = \id_{\k} & (\k \to \k) & \varepsilon[e(k)] = k
\end{array}
\end{eqnarray*}

\medskip

\noindent
{\bf Hopf algebra:} add to the previous axioms:
\begin{eqnarray*}
\begin{array}{cccc}
(H) & p = m(\id_{\H} \otimes S)\Delta = m(S \otimes \id_{\H})\Delta& (\H \to \H) & a_{(1)}(Sa_{(2)}) = (Sa_{(1)})a_{(2)} \\
&  = e\varepsilon && = (\varepsilon a){\mathbf 1}\\
& S*\id = \id* S = e\varepsilon
\end{array}
\end{eqnarray*}
\end{def-notation}

\begin{proposition}\label{pA.1}
Let ${\H}(\cdot, {\Delta}, e,{\varepsilon},S)$ be 
a Hopf algebra over ${\k}$. One has the following properties:
\begin{eqnarray*}
\begin{array}{lcccc}
(i) & (Hm) & Sm = mP_{12}(S \otimes S) & \mbox{i.e.} & S(ab) = (Sb)(Sa),\ a,b, \in H,\\
(ii) & (H\Delta) & \Delta S = (S \otimes S)P_{12}\Delta & \mbox{i.e.} & \Delta Sa = Sa_{(2)} \otimes Sa_{(1)},\ a \in H,\\
(iii) && \left\{\begin{array}{ll}
(He) & S{\mathbf 1} = {\mathbf 1}\\
(H\varepsilon) & \varepsilon S = \varepsilon\\
(Hp) & Sp = pS = p
\end{array}\right. .
\end{array}
\end{eqnarray*}
\end{proposition}

\begin{proof}[Proof of Proposition \ref{pA.1}]
(i) resp. (ii) see below \ref{B.3} resp. {B.4}. 
(iii) $(He)$: by  $(H)$, $(Be)$ and $(B{\varepsilon}e)$:
$$m(S{\otimes}\id_{\H})({\Delta}{\mathbf 1} = {\mathbf 1} {\otimes}{\mathbf 1})=S{\mathbf 1}=(e{\varepsilon}){\mathbf 1} =({\varepsilon}{\mathbf 1}){\mathbf 1}={\mathbf 1}.$$
$(Hp)$:  for $a{\in}{\H}$: 
$$Spa=S[({\varepsilon}a){\mathbf 1}]=({\varepsilon}a)S{\mathbf 1}=({\varepsilon}a){\mathbf 1}=pa,$$ 
moreover, by $(H)$, $(H{\Delta})$, $(Hm)$:
\begin{eqnarray*}
pS&=&m(S{\otimes}\id_{{\H}}){\Delta}S=m(S{\otimes}\id_{{\H}})(S{\otimes}S)
P_{12}{\Delta}\\
&=&mP_{12}(S{\otimes}S)(\id_{{\H}}{\otimes}S)\Delta\\
&=&Sm(\id_{{\H}}{\otimes}S){\Delta}=Sp.
\end{eqnarray*}
$(H{\varepsilon})$:  by $(Hp)$ for $a{\in}{\H}$: 
$$pSa={\varepsilon}(Sa){\mathbf 1}=pa={\varepsilon}(a){\mathbf 1},$$ 
hence ${\varepsilon}(Sa)={\varepsilon}(a)$.
\end{proof}

\begin{proposition}\label{pA.2}
 (The short exact sequence of  ${\varepsilon}$.)

(i)\ The short exact sequence of ${\varepsilon}$ is split 
with lift $e$:
\begin{eqnarray}\label{A.1}
0 \to {\mathbf H}^+ = \mbox{ Ker}\varepsilon \ 
\hookrightarrow {\mathbf H}
\begin{array}{c}
\stackrel{\varepsilon}{\rightarrow\hspace{-0.06cm}\mid}\\
\stackrel{\hookleftarrow}{e} \\
\end{array} \k \to 0
\end{eqnarray}
(where $\H^{+}$ is a notation for Ker${\varepsilon}$) corresponding 
to the following direct sum splitting into a ``vertical" and a ``horizontal" part:
\begin{eqnarray}\label{A.2}
\H = {\mathbf H}^+ \oplus \k {\mathbf 1}.
\end{eqnarray}
The splitting of short exact sequence (1,1) can alternatively 
be specified by:

-- the ``horizontal" projection  $p=e{\varepsilon}$

-- the ``vertical" projection  $p^{+}=\id_{\H}-p=\id_{\H}-e{\varepsilon}$ 
(with Imp$_{+}={\H}^{+} = $ Ker${\varepsilon}$).
\end{proposition}

\begin{proof}[Proof of Proposition \ref{pA.2}]

 (i)\  one has ${\varepsilon}e=\id_{\k}$, 
indeed, for ${\lambda}{\in}{\k}$: $${\varepsilon}e{\lambda}{=}{\varepsilon}({\lambda}{\mathbf 1}){=}{\lambda}{\varepsilon}({\mathbf 1})={\lambda}.$$ The 
other claims follow from known facts about split short exact 
sequences.
\end{proof}

\section{Convolution} 

 In what follows the ground field is ${\k}$ (with possible choices 
$\C, \R, \Q$). We write ${\otimes}$, Hom, and End for ${\otimes}_{{\k}}$, 
Hom$_{{\k}}$, and End$_{{\k}}$.

\begin{def-proposition}\label{defB.1}
With ${\bC}({\Delta},{\varepsilon})$ a coalgebra, ${\A}(m,e)$ 
an algebra; and Hom$(\bC,{\A})$ the vector space of 
homomomorphisms: ${\bC}{\rightarrow}{\A}$ (as ${\k}$-vector spaces), 
define as follows $\Phi * \Phi' \in $ Hom$(\bC,\A)$: 
\begin{eqnarray}\label{B.1}
\Phi * \Phi' = m(\Phi \otimes \Phi')\Delta, \qquad \Phi,\Phi' \in \mbox{ Hom}(\bC,\A).
\end{eqnarray}

Under this product called {\bf convolution} and with $e{\varepsilon}$ as 
a unit, Hom$(\bC,{\A})$ becomes an algebra. Specifically 
one has :
\begin{eqnarray}\label{B.2}
\Phi *(\Phi' * \Phi^{\prime\prime}) = (\Phi * \Phi') * \Phi^{\prime\prime} &=& m^{(2)}(\Phi \otimes \Phi' \otimes \Phi^{\prime\prime})\Delta^{(2)}, \nonumber\\ && \Phi,\Phi',\Phi^{\prime\prime} \in \mbox{ Hom}(\bC,\A)
\end{eqnarray}
(more generally
\begin{eqnarray}\label{B.3}
\Phi_1 * \cdots * \Phi_{n+1} &=& m^{(n)}(\Phi_1 \otimes \cdots \otimes \Phi_{n+1})\Delta^{(n)}, \nonumber\\
&& \Phi_1,\ldots,\Phi_{n+1} \in \mbox{ Hom}(\bC,\A),\ n \in \N),
\end{eqnarray}
in Sweedlers' notation:
\begin{eqnarray}\label{B.1a}
(\Phi * \Phi')a = (\Phi a_{(1)})(\Phi'a_{(2)}), \quad \Phi,\Phi' \in \mbox{ Hom}(\bC,\A),\ a \in \A,
\end{eqnarray}
(more generally
\begin{eqnarray}\label{B.3a}
(\Phi_1 * \cdots * \Phi^{\prime\prime}_{n+1})a &=& m^{(n)}(\Phi_1 a_{(1)})  \cdots (\Phi_{n+1}a_{(n+1)}), \nonumber\\
 \Phi_1,\ldots,\Phi_{n+1} \in \mbox{ Hom}(\bC,\A),\ a \in \A),
\end{eqnarray}
finally
\begin{eqnarray}\label{B.4}
\Phi * e\varepsilon = e\varepsilon * \Phi = \Phi, \qquad \Phi \in \mbox{ Hom}(\bC,\A),\ a \in \A.
\end{eqnarray}
\end{def-proposition}

\begin{proof}
We have  $\Phi * \Phi' \in \mbox{ Hom}(\bC,\A)$ because $\Delta \in \mbox{ Hom}(\bC,\bC \otimes \bC)$,
$\Phi \otimes \Phi' \in  \mbox{ Hom}(\bC \otimes \bC,\bC \otimes \bC)$, $m \in \mbox{ Hom}(\bC \otimes \bC, \bC)$. Check of (\ref{B.2}), we have:
\begin{eqnarray*}
(\Phi * \Phi') * \Phi^{\prime\prime} &=& m(m((\Phi \otimes \Phi')\Delta)\otimes \Phi^{\prime\prime})\Delta\\
&=& m(m \otimes \id)((\Phi \otimes \Phi')\otimes \Phi^{\prime\prime})(\Delta \otimes \id)\Delta\\
&=& m^{(2)}(\Phi \otimes \Phi' \otimes \Phi^{\prime\prime})\Delta^{(2)},
\end{eqnarray*}
whilst
\begin{eqnarray*}
\Phi*(\Phi' * \Phi^{\prime\prime}) &=& m(\Phi \otimes (m(\Phi \otimes \Phi')\Delta)\Delta \\
&=& m(\id \otimes m)(\Phi \otimes (\Phi' \otimes \Phi^{\prime\prime}))(\id \otimes \Delta)\Delta.
\end{eqnarray*}

Check of (\ref{B.1a}), we have: 
\begin{eqnarray*}
(\Phi * \Phi')a &=& m(\Phi \otimes \Phi')\Delta a = m(\Phi \otimes \Phi')
 (a_{(1)} \otimes a_{(2)})\\
&=& m[(\Phi a_{(1)})\otimes (\Phi'a_{(2)})]\\
&=& (\Phi a_{(1)}) \otimes (\Phi'a_{(2)}).
\end{eqnarray*}

Check of (\ref{B.4}),
$$(\Phi * e\varepsilon)a = (\Phi'a_{(1)})(\varepsilon(a_{(2)})){\mathbf 1} = \Phi[a_{(1)}\varepsilon(a_{(2)})]{\mathbf 1} = (\Phi a){\mathbf 1} = \Phi a,$$
and 
$$(e\varepsilon * \Phi)a = (\varepsilon(a_{(1)}){\mathbf 1})(\Phi a_{(2)}) = \Phi[\varepsilon(a_{(1)})a_{(2)}]{\mathbf 1} = (\Phi a){\mathbf 1} = \Phi a, \quad a \in \A.$$
\end{proof}

\begin{example}
 (i)\  Result \ref{defB.1} applies 
in particular to the case where ${\bC}={\A}(m,e,{\Delta},{\varepsilon})$ 
is a bialgebra (with unit element ${\mathbf 1}$). Hom$({\A},{\A})=\mbox{ End}({\A})$ 
then possesses an algebra structures, with product $*$ for which 
$p=e{\varepsilon}$ acting as $p*{\Phi}={\Phi}$, ${\Phi}{\in}\mbox{ End}({\A})$, is 
the unit. 

(ii)\ The further specialization ${\bC}={\A}={\H}$ 
with ${H}(m,e,{\Delta},{\varepsilon},S)$ a Hopf algebra interprets 
the Hopf axiom as the fact that $(H)$: $S*\id_{{\H}}=\id_{{\H}}{*}S=e{\varepsilon}$ 
(this immediately implying uniqueness of the antipode, $S*\id_{{\H}}=e{\varepsilon}$ and 
$\id_{{\H}}*S=e{\varepsilon}$ implying $S*\id_{{\H}}*S'=S=S'$).
\end{example}
\begin{example}
With ${\mathbf H}(m,e,\Delta,\varepsilon,S)$ 
a Hopf algebra, taking for ${\mathbf C}$ the tensor product
coalgebra 
${\mathbf H}\otimes{\mathbf H}(\Delta_{\otimes},\varepsilon_{\otimes})$
and for ${\mathbf A}$ the algebra ${\mathbf H}(m,e)$, we get 
for Hom$({\mathbf H}\otimes{\mathbf H},{\mathbf H})$
an algebra-structure with product $*$ and unit $e\varepsilon_{\otimes}$.
For the latter $Sm$ is a left, and $mP_{12}(S\otimes S)$
is a right inverse of $m: Sm*m=e\varepsilon_{\otimes}=m*mP_{12}(S\otimes S)$
 (they thus coincide, this proving property (Hm)).
\end{example}
\begin{example}
With ${\mathbf H}(m,e,\Delta,\varepsilon,S)$ 
a Hopf algebra, taking for ${\mathbf C}$  the coalgebra 
${\mathbf H}(\Delta,\varepsilon)$ and for ${\mathbf A}$ the 
tensor product algebra
${\mathbf H}\otimes{\mathbf H}(m_{\otimes},\varepsilon_{\otimes})$
we get for 
Hom$({\mathbf H},{\mathbf H}\otimes{\mathbf H})$
an algebra structure with product $*$ and unit $e_{\otimes}\varepsilon$.
For the latter $\Delta S$ is a left, and 
$ \ (S\otimes S)P_{12}\Delta$  is a right inverse of 
$\Delta: \ \Delta S*\Delta = e_{\otimes}\varepsilon = 
 \Delta*(S\otimes S)P_{12}\Delta \ $
(they thus coincide, this proving property (H$\Delta$)).
\end{example}

\section{Primitive and group-like elements}

\begin{definition}\label{defC.1}
 Let ${\H}({\Delta},{\varepsilon})$ be a bialgebra over the field ${\k}$.

(i)\ The element $a{\in}{\H}$ is called {\bf primitive} whenever
\begin{eqnarray}\label{C.1}
\Delta a = a \otimes {\mathbf 1} + {\mathbf 1} \otimes a.
\end{eqnarray}
The set of primitive elements of ${\H}$ is denoted by $P({\H})$.

(ii)\ The element $a{\in}{\H}$ is called {\bf group-like} whenever
\begin{eqnarray}\label{C.2}
a \neq 0 \qquad \mbox{ and } \qquad \Delta a = a \otimes a.
\end{eqnarray}
The set of group-like elements of ${\H}$ is denoted by $G({\H})$.
\end{definition}

\begin{proposition}
 Let ${\H}({\Delta},{\varepsilon})$ be a ${\k}$-bialgebra.

(i)\ $P({\H})$ is a ${\k}$-linear subspace of ${\H}$. Moreover 
if $a,b{\in}P({\H})$, then $[a,b]=ab-ba{\in}P({\H})$.
Consequently $P({\H})$ is a Lie subalgebra of the Lie algebra Lie${\H}$. 

(ii)\ One has  $P({\H}){\subset} \mbox{ Ker}{\varepsilon}$:
\begin{eqnarray}\label{C.3}
a \in P(\H)\Rightarrow \varepsilon(a) = 0,
\end{eqnarray}

(iii)\ $G({\H})$ is multiplicative: if $a,b{\in}G({\H})$,  then $ab{\in}G({\H})$.

(iv)\ One has
\begin{eqnarray}\label{C.4}
a \in G({\mathbf H})  \Rightarrow \varepsilon(a) = 1.
\end{eqnarray}

If furthermore ${\H}$ is a Hopf algebra with antipode $S$, then:
\begin{eqnarray}
a \in P(\H) &\Rightarrow& Sa = -a.\label{C.5}\\
a \in G(\H) &\Rightarrow& Sa = a^{-1},\label{C.6}
\end{eqnarray}
showing that $G({\H})$ is then a group.
\end{proposition}

\begin{proof}
 (i)\ Let $a,b{\in}P({\H})$, ${\lambda}{\in}{\k}$, 
then $a+b,{\lambda}a{\in}P({\H})$. Furthermore:
\begin{eqnarray*}
\Delta [a,b] &=& \Delta(ab - ba) = (\Delta a)(\Delta b) - (\Delta b)(\Delta a)\\
&=& (a \otimes {\mathbf 1} + {\mathbf 1} \otimes a)(b \otimes {\mathbf 1} + {\mathbf 1} \otimes b) - (b \otimes {\mathbf 1} + {\mathbf 1} \otimes b)(a \otimes {\mathbf 1} + {\mathbf 1}\otimes a)\\
&=& ab \otimes {\mathbf 1} + a \otimes b + b \otimes a + {\mathbf 1} \otimes ab - (ba \otimes {\mathbf 1} + b \otimes a + a \otimes b + {\mathbf 1} \otimes ba)\\
&=& (ab - ba) \otimes {\mathbf 1} + {\mathbf 1} \otimes (ab -ba) = [a,b] \otimes {\mathbf 1} + {\mathbf 1} [a,b].
\end{eqnarray*}

(ii)\  $(\id{\otimes}{\varepsilon}){\Delta}{=}\id$ entails 
\begin{eqnarray*}
(\id \otimes \varepsilon)\Delta a & = & (\id \otimes \varepsilon)
 (a \otimes {\mathbf 1} + {\mathbf 1} \otimes a )  \\ 
  & = &   a \otimes {\mathbf 1} + {\mathbf 1}\otimes\varepsilon(a) \\
  & = & a + \varepsilon(a){\mathbf 1}  \ = \ a.
\end{eqnarray*}

(iii)\ $a,b{\in}G({\H})$ entails 
$$\Delta (ab) = (\Delta a)(\Delta b) = (a \otimes a)(b \otimes b) = ab \otimes ab.$$

(iv)\ $(\id{\otimes}{\varepsilon}){\Delta}{=}\id$ entails 
$$(\id \otimes \varepsilon) \Delta a = (\id \otimes \varepsilon)(a \otimes a) = a \otimes \varepsilon(a) = a \varepsilon(a) = a.
$$
Assume that ${\H}$ is a Hopf algebra with antipode $S$. 

Check of (\ref{C.5}): $m(\id{\otimes}S){\Delta}=e{\varepsilon}$ entails 
$$0=m(\id{\otimes}S)(a{\otimes}{\mathbf 1}+{\mathbf 1}{\otimes}a)=m(a{\otimes}{\mathbf 1}+{\mathbf 1}{\otimes}Sa)=a+Sa.$$

Check of (\ref{C.6}): $m(\id{\otimes}S){\Delta}=e{\varepsilon}$ entails 
$$m(\id \otimes S)(a \otimes a) = m(a \otimes Sa) = a(Sa) = {\mathbf 1}.$$
whilst $m(S{\otimes}\id){\Delta}=e{\varepsilon}$ entails $$m(S{\otimes}\id)(a{\otimes}a)=m(Sa{\otimes}a)=(Sa)a={\mathbf 1}.$$
\end{proof}


\begin{thebibliography}{}
\bibitem{Connes1} A. Connes and D. Kreimer. Renormalization in quantum field theory 
and the Riemann-Hilbert Problem. I: the Hopf algebra structure 
of graphs and the main theorem.
hep-th/9912092 13 Dec. 1999. 

\bibitem{Connes2} A. Connes and D. Kreimer. Renormalization in quantum field theory 
and the Riemann-Hilbert Problem. II: the ${\beta}$-function, 
diffeomorphisms and the renormalization  group. arXiv: hep-th/0003 
188 21 Mars 2000.

\bibitem{B}  N. Bourbaki. Alg\`{e}bre. Chapitre II. Par. 6.
\end{thebibliography}
\end{document}